\documentclass[iop]{emulateapj}

\usepackage{amsmath, amsthm}
\usepackage{booktabs}

\slugcomment{Submitted, July 25, 2016}

\newcommand*{\Lx}{L_{\rm X}}
\newcommand*{\Tx}{T_{\rm X}}

\newcommand*{\Mstar}{M_{\star}}
\newcommand*{\DMstar}{\Delta\Mstar}
\newcommand*{\Mdotstar}{{\dot M}_{\star}}

\newcommand*{\Mbh}{M_{\rm BH}}
\newcommand*{\DMbh}{\Delta\Mbh}
\newcommand*{\Mdotbh}{{\dot M}_{\rm BH}}
\newcommand*{\Mbhz}{M_{\rm BH,0}}

\newcommand*{\Mdotout}{{\dot M}_{\rm out}}
\newcommand*{\Mdotin}{{\dot M}_{\rm in}}

\newcommand*{\Mdotedd}{{\dot M}_{\rm Edd}}

\newcommand*{\mdot}{\dot{m}}

\newcommand*{\Mh}{M_{\rm h}}

\newcommand*{\Mdotsn}{{\dot M}_{\rm SN}}
\newcommand*{\Rsn}{R_{\rm SN}}
\newcommand*{\Esn}{E_{\rm SN}}

\newcommand*{\Ledd}{L_{\rm Edd}}
\newcommand*{\Lsn}{L_{\rm SN}}
\newcommand*{\Lbh}{L_{\rm BH}}
\newcommand*{\Lb}{L_{\rm B}}

\newcommand*{\Lw}{L_{\rm w}}
\newcommand*{\vw}{v_{\rm w}}
\newcommand*{\vwz}{v_{\rm w0}}
\newcommand*{\pdotw}{{\dot p}_{\rm w}}
\newcommand*{\rhodotw}{{\dot\rho}_{\rm w}}
\newcommand*{\epsw}{\epsilon_{\rm w}}
\newcommand*{\epswz}{\epsilon_{\rm w0}}
\newcommand*{\Aw}{A_{\rm w}}

\newcommand*{\epsem}{\epsilon_{\rm EM}}
\newcommand*{\epsemz}{\epsilon_{0}}
\newcommand*{\Aem}{A_{\rm EM}}

\newcommand*{\vvstar}{{\bf v}_{\star}}
\newcommand*{\vvw}{{\bf v}_{\rm w}}
\newcommand*{\uv}{{\bf u}}

\newcommand*{\tnow}{t_{12}}
\newcommand*{\Msun}{M_{\odot}}

\newcommand*{\Tr}{{\rm Tr}}
\newcommand*{\gradient}{\nabla}
\newcommand*{\dtpartial}[1]{\dfrac{\partial#1}{\partial t}}
\newcommand*{\norm}[1]{\left\lVert#1 \right\rVert}
\newcommand*{\reo}{R_{\rm e\,0}}
\newcommand*{\re}{R_{\rm e}}
\newcommand*{\se}{\sigma_{\rm e8}}
\newcommand*{\kms}{\rm{km~s}^{-1}}
\newcommand*{\ergs}{\rm{erg~s^{-1}}}
\newcommand*{\fDM}{f_{\rm DM}}
\newcommand*{\rhoh}{\rho_{\rm h}}
\newcommand*{\rh}{r_{\rm h}}
\newcommand*{\rhocrit}{\rho_{\rm crit}}
\newcommand*{\deltac}{\delta_{\rm c}}
\newcommand*{\Tc}{T_\mathrm{c}}
\newcommand*{\lbhefphot}{L_{\rm BH,photo}^{\rm eff}}

\newcommand*{\vphi}{v_{\varphi}}

\shortauthors{L. Ciotti et al.}
\shorttitle{2D hydrodynamical simulations of the BAL wind feedback}

\begin{document}

\title{The Effect of the AGN Feedback on the Interstellar Medium of Early-Type Galaxies:
2D Hydrodynamical Simulations of the Low-Rotation Case.}
\author{Luca Ciotti\altaffilmark{1}, Silvia Pellegrini\altaffilmark{1}, Andrea Negri\altaffilmark{2}, Jeremiah P. Ostriker\altaffilmark{3,4}}
\affil{$^1$Department of Physics and Astronomy, University of Bologna, via Ranzani 1, I-40127, Bologna, Italy} 
\affil{$^2$CNRS, UMR 7095, Institut d'Astrophysique de Paris, 98bis bvd Arago, F-75014 Paris, France}
\affil{$^3$Department of Astronomy, Columbia University, 550 W. 120th Street, New York, NY 10027, USA}
\affil{$^4$Princeton University Observatory, Princeton, NJ 08544, USA}

\begin{abstract}

  We present 2D hydrodynamical simulations for the evolution of hot
  gas flows in early-type galaxies with central massive black holes
  (MBHs), starting from an age of $\approx 2$ Gyr; the code has an
  accurate and physically consistent description of radiative and
  mechanical (due to AGN winds) feedback, and a parsec-scale
  resolution at the center. The mass input for the flow comes from
  stellar mass losses, and the energy input includes Type Ia supernova
  and stellar heating; the flow can form stars.  Realistic,
  axisymmetric dynamical models for the galaxies are built by solving
  the Jeans' equations.  We find that the lowest mass models explored
  ($\Mstar=8\times 10^{10}\Msun$) develop a global outflow sustained
  by SNIa’s heating after a few Gyr, and then end with a significantly
  lower amount of hot gas and new stars ($\DMstar$). In more massive
  models, instead, nuclear outbursts last up to the present epoch,
  with large and frequent fluctuations in the emission of the nucleus
  and of the gas ($\Lx$).  Each burst lasts $\sim$few$\times 10^7$ yr,
  during which, in the inner $\sim 2-3$ kpc region, cold, inflowing,
  and hot, outflowing gas phases coexist.  The $\Lx-T$ relation for
  the gas reproduces well that of local galaxies.  AGN activity
  determines a positive feedback on star formation.  Roughly half of
  the total mass losses ends recycled into new stars, just $\simeq
  3$\% of it is accreted on the MBH, and most of the remaining part is
  ejected from the galaxy; the ratio between the mass of gas displaced
  outer of 5$\re$ during the evolution and that in new stars
  $\DMstar$, the load factor, is $\simeq 0.6$.  A rounder galaxy shape
  corresponds to a larger final MBH mass, $\DMstar$, and $\Lx$.  Half
  of the radiative output from the AGN is emitted within $\approx 3$
  Gyr from the start of the simulation, roughly the same timescale
  within which half of the new stars form; almost all of the time is
  spent at very low nuclear luminosities, yet one quarter of the total
  energy is emitted at an Eddington ratio $>0.1$.  The duty-cycles of
  AGN activity ranges from 3 to 5\%.

\end{abstract}

\section{Introduction}
 
The relationship of the QSO activity at high $z$ with the surrounding
ISM of the host galaxy, including the possibility of triggering or
quenching star formation, is one of the currently most debated and
still unsettled topics in the field of galaxy evolution.  The massive
black holes (hereafter MBHs) at the centers of massive elliptical
galaxies were in place already when the universe was $\simeq 1$ Gyr old
(e.g., Madau \& Rees 2001, Alvarez et al. 2009, Wu et al. 2015), and
fast, large scale, and massive outflows, driven by QSOs, are supposed
to transform young, star-forming galaxies into ``red and dead''
spheroids (Sturm et al. 2011, Cano-D\'iaz et al. 2012,
Faucher-Gigu\`ere \& Quataert 2012, Feruglio et al. 2015, King \&
Pounds 2015).  But the QSO and starformation activities are not likely
to suddenly stop at high $z$.  Concerning the origin of the QSO
activity, it appears that the most luminous AGN phases, preferentially
found at $z>2$, may be connected to direct accretion of cold gas and
to mergers (e.g., Di Matteo et al. 2005, Kazantzidis et al. 2005,
Dubois et al. 2012), while less luminous and lower $z$ AGNs seem to be
driven by other processes, unrelated with the merging phenomenon, in
general (Cisternas et al. 2011, Treister et al. 2012, Schawinski et
al. 2012, Kocevski et al. 2012; see Heckman \& Best 2014 for a
review).

A likely possibility is that the stellar mass losses normally produced
during stellar evolution cyclically feed a central gas inflow (Norman
\& Scoville 1988), and then trigger the QSO activity also for isolated
early-type galaxies (hereafter ETGs), that are often considered
``dead''.  Indeed, these losses represent a major source in mass for
the ISM (e.g., Ciotti et al. 1991); also, the high metallicity of the
observed outflows from low and high-$z$ galaxies provides evidence
that the fuel source for these flows is highly processed gas, not
"cold flows" accreted from outside (Cooksey et al. 2010, Fox 2011,
Lehner et al. 2013). Numerical simulations describing the evolution of
stellar mass losses in ETGs provided clear support for the conjecture
that they trigger the QSO activity at epochs closer than $z\sim 2$
(Ciotti et al. 2010). The galaxy-black hole coevolution under the
effects of these internal processes has been generally termed
``secular evolution'' (e.g., Heckman \& Best 2014).

In a number of previous works, Ciotti, Ostriker and coworkers
investigated the relationship between the secular evolution of the
stellar population in ETGs, the QSO activity, and the resulting
feedback action on the host galaxy, since $z\sim 2$, {\it after} the
MBH and the stellar population have terminated their major growing
phase\footnote{For a standard cosmology $z=2$ corresponds to an age of
  the universe of 3.3 Gyr, an epoch at which massive ETGs are thought
  to have completed the bulk of their star formation process, and MBH
  formation.}.  With high-resolution hydrodynamical simulations in
spherical symmetry, they studied in detail the mechanical and
radiative feedback effects induced by accretion of stellar mass
losses. The physics of feedback was modeled on the observed dominant
processes: radiative output and BAL wind output. In fact, the emitted
photons impart energy and momentum to the ISM via electron scattering,
photoionization, scattering due to atomic resonance lines, and
absorption by dust grains; in addition, accretion drives broad
absorption line (BAL) winds that convey mass, momentum, and energy to
the ISM surrounding the nucleus (as expected: Silk \& Rees 1998, King
2003, Bieri et al. 2016; and as intensively observed: Reichard et
al. 2003, Greene et al. 2011, Arav et al. 2013, Liu et al. 2013,
Carniani et al. 2015, McElroy et al. 2015).  The simulations
considered the corresponding cooling and heating functions, including
photoionization plus Compton scattering, and solved the radiative
transport equations, also in presence of dust; they allowed for mass
and energy inputs from stellar winds, and Type Ia and Type II
supernovae (hereafter respectively SNIa and SNII); and they considered
the mechanical feedback due to the nuclear wind and star formation
(hereafter SF) induced by accretion.

These simulations covered length scales from $\sim 5$ pc to $\sim 200$
kpc, and timescales from $\sim 10^2$ yr (or less) to $10^{10}$ yr;
thus, all the relevant length and time-scales were resolved (from the
Bondi accretion radius to tens of optical effective radii), and the
accretion rates, as well as the effect of AGN feedback on the gas over
the whole galaxy, were self-consistently determined.  This, together
with the complex but exhaustively implemented input physics, was a
specific and very important feature of the simulations, crucial to
establish what is the exact accretion rate, and then the feedback
effects on the final MBH mass, on the ISM, and on SF. In other
numerical studies (e.g., Di Matteo et al. 2005, 2008; Booth \& Schaye
2009; Choi et al. 2015; Sijacki et al. 2015) the mass accretion rate
remains ``unresolved'', and is set from recipes or algorithms, due to
the lack of all the required spatial extent and resolution, or to
difficulties in resolving the gas mass distribution intrinsic to the
numerical modeling (e.g., in the SPH codes).  For example,
cosmological simulations suffer inevitable limits due to numerical
resolution, and the mass accretion rate is given by recipes generally
based on the Bondi rate (Bondi 1952), or ad hoc prescriptions based on
it (with all the associated uncertainties; e.g. Curtis \& Sijacki
2016, Korol et al. 2016).  Also, the effect of the MBH on its
surroundings has often been modeled by injecting thermal energy into
the ISM again following recipes for its amount and distribution
(typically with the goal of reproducing observed relations or general
properties).

The simulations in spherical symmetry of Ciotti, Ostriker and
coworkers showed that in the medium-high mass ETGs the resulting
evolution is highly unsteady (e.g., Ciotti, Ostriker \& Proga 2010).
At early times (starting from $z\sim 2$) major accretion episodes
caused by cooling flows trigger AGN flaring, with duty cycles small
enough to account for the small fraction of massive galaxies observed
to be in the QSO phase, when the accretion luminosity approaches the
Eddington luminosity. At low redshift, the majority of models are
characterized by smooth, very sub-Eddington mass accretion rates. At
the end of the evolution, the mass of the MBH is limited to the range
of masses observed today, even though the mass lost by the stars is
roughly two orders of magnitude larger than the MBH masses observed in
local ETGs. Note that the MBH heating alone has been shown to be not
sufficient by itself to avoid long-lasting and massive inflows towards
the galactic center at early times, but when coupled with the SNIa's
heating, it becomes very efficient in sustaining the galaxy degassing
and preventing large mass accumulation in the central regions.  During
the evolution roughly half of the mass lost by stars gets ejected in
SNIa driven winds, and half falls to the center and ends accreted or
in starbursts.  This series of simulations in spherical symmetry was
followed by an investigation in two dimensions, again applied to
spherical galaxies (Novak et al. 2011, 2012; Gan et al. 2014). A 2D
implementation of the same feedback physics described above showed
that MBH accretes some of the infalling gas and expels a conical wind;
and that the cool shells, forming at 0.1$-$1 kpc from the center, are
Rayleigh–Taylor unstable to fragmentation, leading to a somewhat
higher accretion rate, and less effective feedback.

In the current work we study the radiative and mechanical (due to AGN
winds) feedback effects on the ISM by improving the treatment of our
previous 2D works in two main respects: 1) the galaxy models
underlying the ISM evolution are more realistic and accurate than ever
previously: they are axisymmetric, allow for various degrees of
flattening, and include, in addition to the MBH, a generalized de
Vaucouleurs stellar profile coupled with a NFW (Navarro et al. 1997)
dark matter halo; the halo contributes less dark mass than the stellar
mass within one effective radius, as observed recently for ETGs (e.g.,
Cappellari et al. 2015); the stellar kinematics is determined solving
the Jeans equations for the total mass (MBH+stars+dark halo) and a
chosen orbital distribution; all galaxy parameters are determined in
order to keep the models on the main observed scaling laws. 2) We
consider a secularly evolving stellar population input, i.e., a
secularly decreasing stellar mass loss rate and SNIa explosion rate,
distributed over the galaxy.  With these two improvements implemented
in our 2D simulations with well resolved length and time scales, we
have addressed the following questions: what is the effect of feedback
on the MBH mass?  What is the increase of the MBH mass due to
accretion at epochs more recent than $z\sim 2$?  Is it plausible that
the ${\rm \Mbh}-\sigma$ relation was already in place at $z\sim 2$?
What is the effect of radiative feedback and BAL winds on the ISM?
Are mass losses and AGN activity connected with residual SF episodes
at late times (i.e., after most of the SF has completed)?  Is feedback
responsible for more or less SF? Do we correctly predict the
properties of the circumgalactic medium?

The paper is organized a follows: Sect. 2 describes the galaxy models;
Sect. 3 presents the numerical code and the hydrodynamical equations
it solves, the inputs to them, the implementation of the feedback and
SF physics; Sect. 4 presents the results of the simulations; Sect. 5
summarizes the main conclusions.

\section{Galaxy models}
We briefly summarize here the main characteristics of the galaxy
models, with respect to their stellar population properties and
evolution (Sect. 2.1), and their internal dynamics (Sect. 2.2).

\subsection{Stellar population}
In ETGs the gas is lost by evolved
stars mainly during the red giant, asymptotic giant branch, and
planetary nebula phases. These losses originate ejecta that initially
have the velocity of the parent star, then interact with the mass lost
from other stars or with the hot ISM, and mix with it (Mathews 1990,
Parriott \& Bregman 2008, Bregman \& Parriott (2009).  
Thus, stellar winds are heated to X-ray temperatures by thermalization of the kinetic energy of 
collisions between stellar ejecta, as will be presented in Sect. 2.2 below.
Far infrared observations allow for measurements 
of the stellar mass loss rate for the whole galaxy ($\Mdotstar$), giving an average rate in
reasonable agreement with theoretical predictions (Athey et al. 2002).
According to single burst stellar population synthesis models (Maraston 2005), the trend of $\Mdotstar$ with time, 
for solar metal abundance, after an age of $\ga 2$ Gyr, can be approximated as:
\begin{equation}
\Mdotstar (t) = 10^{-12}\, A\,\times \Mstar \,\, \tnow^{-1.3} 
\quad\quad\quad ({\rm\Msun yr^{-1}}),
\label{eq:mdots}
\end{equation}
where $\Mstar$ is the galactic stellar mass in solar masses
at an age of 12 Gyr, $\tnow$ is the age in units of 12 Gyrs, and $A=2.0$ or 3.3 for a
Salpeter or Kroupa IMF (the latter is adopted here; see also Pellegrini 2012).
The relation above agrees well with previous theoretical estimates
(e.g., Mathews 1989, Ciotti et al. 1991).  


Also SNIa's explosions provide mass and heat to the ISM, and the total
mass loss rate of a stellar population is ${\dot M}(t)=\Mdotstar (t)+
\Mdotsn (t)$, where the mass input due to SNIa's is $\Mdotsn
(t)=1.4\Msun\, \Rsn (t)$. Here $\Rsn (t)$ (in yr$^{-1}$) is the
evolution of the explosion rate with time, and each SNIa ejects
$1.4\Msun$. In models of SNIa's explosions past a burst of SF (Greggio
2010), $\Rsn (t)$ experiences a raising epoch during the first
$\simeq 1$ Gyr, and then decreases slowly with a timescale of the
order of 10 Gyr, down to the present day observed rate.  A
parameterization of the rate after the peak, in number of events per
year, is
\begin{equation}
\Rsn (t)=0.16  (H_0/70)^2 \times 10^{-12} \,\,
\Lb \,\,\, \tnow^{-s}\quad\quad ({\rm yr}^{-1}) ,
\label{eq:rsn}
\end{equation}
where $H_0$ is the Hubble constant in units of km s$^{-1}$ Mpc$^{-1}$,
$\Lb$ is the present epoch B-band galaxy luminosity in $L_{B,\odot}$,
and $s$ characterizes the secular evolution; when $\tnow=1$,
Eq.~\ref{eq:rsn} gives the rate for local ETGs in recent measurements
(e.g. Mannucci et al. 2005, Maoz et al. 2011).  For the rate in
Eq.~\ref{eq:rsn} and $H_0=70$ km s$^{-1}$ Mpc$^{-1}$, one obtains
$\Mdotsn (12 Gyr) =2.2 \times 10^{-13}\,\Lb\Msun$ yr$^{-1}$, that is
almost $\sim 100$ times smaller than the "quiescent" stellar mass loss
rate $\Mdotstar (12 Gyr)\approx 2\times 10^{-11} \, \Lb\Msun$
yr$^{-1}$ given above. Recent estimates of the slope $s$ agree with a
value around $s\simeq 1$ (Maoz et al. 2011, Sharon et al. 2010).

The heating rate provided by SNIa's explosions $\Lsn(t)$ is the
product of the kinetic energy injected by one event ($\Esn\approx
10^{51}$ erg) times the rate $\Rsn (t)$, and times an efficiency
factor.  In the code we adopt an efficiency of 0.85, an assumption
that is not unreasonable for the hot diluted gas (see also Sect. 3.1).
Then at most $\Lsn(t) = \Esn\Rsn (t)$, that is:
\begin{equation}
\Lsn(t) =5.1  (H_0/70)^2 \times 10^{30}
\Lb \,\, \tnow^{-s}\quad\quad ({\rm erg \,\, s}^{-1}).
\label{eq:lsn}
\end{equation}

Of course, another major source of ISM heating is provided by the central
MBH, as will be discussed in Sect. 3 below.

\subsection{Dynamical structure}

We consider here a subset of the large suite of models built for the
study of gas flows in galaxies of various shapes and internal
kinematics of Negri et al. (2014b).  These are axisymmetric galaxy
models, where, in addition to the central MBH (of initial mass
$\Mbhz=10^{-3}\Mstar$), there is a stellar component characterized by
different intrinsic flattenings, and a spherical dark matter (DM)
halo. The stellar density is described by the ellipsoidal deprojection
(Mellier \& Mathez 1987) of the de Vaucouleurs (1948) law:
\begin{equation}
\rho_{\star}(R,z)=\rho_0\zeta ^{-0.855}\exp(-\zeta ^{1/4}),
\label{eq:rho_*}
\end{equation}
with
\begin{equation}
\rho_0=\dfrac{\Mstar b^{12}}{16\pi q\reo^3\Gamma(8.58)},\quad
\zeta =\dfrac{b^4}{\reo}\sqrt{R^2+\dfrac{z^2}{q^2}},
\label{eq:rho_*2}
\end{equation}
where $(R,\varphi,z)$ are the cylindrical coordinates,
$b\simeq7.67$, $\reo$ is the projected half mass radius (effective radius) when
the galaxy is seen face-on\footnote{For an edge-on view, the circularized
effective radius is $\re=\reo\sqrt{q}$.}, and the parameter
$q\leqslant 1$ controls the flattening, so that the minor axis is aligned with the $z$
axis. For the simulations we
consider $q$ values of (1, 0.6, 0.3), corresponding to E0, E4 and E7
galaxies when seen edge-on. For the DM halo we adopt an untruncated
NFW (Navarro et al. 1997) profile:
\begin{equation} 
\rhoh (r)=\dfrac{\rhocrit~\deltac \rh} {r (1+r/\rh)^2},
\label{eq:NFW}
\end{equation}
where $\rhocrit=3H^2/8\pi G$ is the critical density for closure, and 
\begin{equation}   
\deltac=\dfrac{200}{3} \dfrac{c^3}{\ln(1+c)-c/(1+c)},\quad c\equiv
\dfrac{r_{200}}{\rh}
\end{equation}
and $r_{200}$ is the radius of a sphere of mean interior density of 200$\rhocrit$.
We refer to the DM mass enclosed within $r_{200}$ as to the halo mass $\Mh$. 

All the relevant dynamical properties of the models were computed with
a code built for this purpose (Posacki et al. 2013).  Starting from an
axisymmetric density distribution $\rho_{\star}(R,z)$ produced by a
two-integral phase-space distribution function, the code solves the
Jeans equations in cylindrical coordinates, and computes the velocity
fields of the stars, the total potential $\Phi_{\mathrm{tot}}(R,z)$
due to all components (stars, dark halo, MBH), and the vertical and
radial forces.  The radial and vertical velocity dispersions are equal
($\sigma_R=\sigma_z\equiv\sigma$), and the only non-zero streaming
motion is in the azimuthal direction ($\overline{\vphi}$). To set the
latter, we adopted the Satoh (1980) $k$-decomposition
$\overline{\vphi}^2=k^2(\overline{\vphi^2}-\sigma^2)$, from which the
azimuthal velocity dispersion is recovered as
$\sigma_{\varphi}^2\equiv\overline{\vphi^2}-\overline{\vphi}
^2=\sigma^2+(1-k^2)(\overline{\vphi^2}-\sigma^2)$, where $0\leqslant k
\leqslant 1$. For $k=1$ the galaxy is an isotropic rotator, while for
$k=0$ no net rotation is present, and all the flattening is due to
$\sigma_{\varphi}$. In general, $k$ can be a function of $(R,z)$, and
more complicated (realistic) velocity fields can be realized (Ciotti
\& Pellegrini 1996; Negri et al. 2014a,b).  In any case, $k$ is
bounded from above by the function $k_{max}^2
(R,z)=\overline{\vphi^2}/(\overline{\vphi^2}-\sigma^2)$.  In this work
we restrict to the $k\simeq 0$ case, and then in practice only stellar
random motions are thermalized (see also Sect. 4 for more discussion
on the role and then the adopted values of the $k$ parameter).

The only free parameter of the stellar distribution is $\se$, the
aperture luminosity-weighted velocity dispersion within $\re/8$, from
which the galaxy luminosity is recovered from the Faber--Jackson
relation, and the size $\re$ from the size--luminosity relation
(Desroches et al. 2007). From the stellar mass-to-light ratio, fixed
at that of a 12 Gyr old stellar population with a Kroupa initial mass
function, the stellar mass $\Mstar$ is derived (see Posacki et
al. 2013 for more details). By assigning a given $\se$, we build a
spherical galaxy, that we call the ``progenitor''.  The free
parameters of the DM halo are determined by the need to reproduce the
assumed $\se$, by fixing $\Mh/\Mstar\simeq 20$ (Behroozi et
al. 2013), and imposing that the DM fraction $\fDM$ within a sphere of
radius $\re$ keeps well below unity (Cappellari et al. 2015).  These
constraints produce $\rh\simeq 2\re$, $22\lesssim c \lesssim 41$, and
$\fDM\simeq 0.6$ for the spherical progenitors.  The flattened
descendants of each progenitor have the same circularized $\re$ as the
spherical progenitor, when seen edge-on, thus they expand with
decreasing $q$ ($\rho_{\star}\propto \sqrt{q}$); their $\se$, as a
consequence, also decreases with respect to that of the progenitor
(while $\Mstar$ and $\Lb$ remain the same).  In the flattening
procedure the DM halo is maintained fixed to that of the progenitor
(see Posacki et al. 2013 for a more comprehensive model
description\footnote{The models considered here belong to the class of
  ``edge-on built'' descendants of Posacki et al. (2013), while the
  latter authors also built the class of ``face-on built''
  descendants, that have the same $\re$ as the spherical progenitor
  when seen face-on.}).

We consider here four spherical progenitors, with $\se$ values of 180,
210, 250 and 300 $\kms$, and corresponding stellar masses $\Mstar$ of
0.81, 1.54, 3.35 and 7.80$\times 10^{11}\Msun$.  The simulations are
run for eight flattened descendants: for each progenitor, they have
intrinsic flattenings with $q=0.6$ (E4 shape) and $q=0.3$ (E7 shape).
Table~\ref{tab:params} lists all the relevant parameters
characterizing these eight galaxy models.

\section{Feedback and star formation in the hydrodynamical equations}

We describe here the hydrodynamical equations that are solved to
evolve the gas flow (Sect. 3.1), and the input physics for them, with
particular regard to the implementation of the SF process (Sect. 3.2)
and feedback effects (Sects. 3.3 and 3.4).

\subsection{The hydrodynamical equations}

The hydrodynamical equations that are numerically integrated in spherical coordinates, by a 
significantly updated version of the Novak et al. (2011) code, are the following:
\begin{gather}
\dtpartial{\rho} + \nabla \cdot (\rho\uv) = \dot{\rho}_\mathrm{Ia} + \dot{\rho}_\mathrm{\star} + \dot{\rho}_\mathrm{II} - \dot{\rho}_\mathrm{SF} + \dot{\rho}_\mathrm{w}
\label{eq:continuity},\\
\begin{split}
\hskip 0.5truecm
\rho\dtpartial{\uv} + \rho 
\left(\uv \cdot \nabla \right) \uv = - \nabla p - \rho 
\gradient \Phi_{\rm tot} - \gradient p_\mathrm{rad} + \quad\quad\quad \\ 
+(\dot{\rho}_\mathrm{Ia} + \dot{\rho}_\mathrm{\star} + \dot{\rho}_\mathrm{II})
(\vvstar - \uv) + \rhodotw (\vvw -\uv)
\label{eq:euler},\\
\end{split}
\\
\dtpartial{E} + \nabla \cdot (E \uv) = 
- p\nabla \cdot \uv  + H - C + \dot{E} 
 - \dot{E}_\mathrm{SF} + {\rhodotw\over 2} \norm{\vvw -\uv}^2
\label{eq:energy}
\end{gather}
where $\rho$, $\uv$, $E$, $p$ are respectively the mass density,
velocity, internal energy density, and pressure of the gas;
$\vvstar$ is the streaming velocity of the stellar component;
$\vvw$ is the AGN wind velocity. $H$ is the
volumetric heating rate due to radiative feedback, and $C$ is the
bolometric cooling rate per unit volume (see Sect. 3.4).  The ISM is a
fully ionized monoatomic gas, with $p = (\gamma -1)E$, $\gamma = 5/3$,
and solar composition ($\mu= 0.62$). The mass density rates
$\dot{\rho}_\mathrm{Ia}$ and $ \dot{\rho}_\mathrm{\star} $ describe
the mass injection from the old stellar population, i.e., from SNIa's
explosions and from normal stars (Sect. 2.1); $\dot{\rho}_\mathrm{SF}$
and $\dot{\rho}_\mathrm{II}$ describe the mass sink due to SF, and the
mass input due to the SNII produced by SF (see Sect. 3.2);
$\dot{E}_\mathrm{SF}$ is the energy sink due to SF (Sect. 3.2).
$\dot{E}$ is the total energy injection rate due to the old and new
stellar populations, produced by the thermalization of the kinetic
energy of SNIa and SNII explosions (${\dot E}_{\mathrm{Ia}}$ and
${\dot E}_{\mathrm{II}}$ respectively), and by the thermalization of
the relative motions between stars and the ISM (existing at the moment
of injection). $\dot{E}$ is then given by:
\begin{equation}
\dot E =  {\dot E}_{\mathrm{Ia}}  +
 {\dot E}_{\mathrm{II}}  + \dfrac{\dot\rho_{\mathrm{Ia}}+\dot\rho_{\star}+ \dot\rho
_{\mathrm{II}}}{2} [\norm{\vvstar - \uv}^2 +\Tr ({\boldsymbol{\sigma}}^2)], \label{edot}
\end{equation}
where ${\dot E}_{\mathrm{Ia}} =
\dot\rho_\mathrm{Ia}\vartheta_\mathrm{SNIa}
E_\mathrm{SN}/(1.4~\Msun)$, with $E_\mathrm{SN}=10^{51}$ erg and
$1.4\Msun$ being respectively the kinetic energy and ejected mass of
one SNIa event, and $\vartheta_\mathrm{SNIa}$ is the thermalization
efficiency, for which we adopt the value of 0.85, as a plausible one
for a low density and hot medium (see Mathews 1989, Tang \& Wang
2005); ${\dot E}_{\mathrm{II}}$ is the SNII energy injection rate, and
is calculated considering the overlap in time of subsequent SF
episodes (as for $\dot{\rho}_\mathrm{II}$ below; see Sect. 3.2, and
Negri et al. 2015).  ${\dot E}_{\mathrm{II}}$ also depends on the
thermalization efficiency, that we take equal to
$\vartheta_\mathrm{SNIa}$.  Finally, $\Tr (\boldsymbol{\sigma}^2)$ is
the trace of the stellar velocity dispersion tensor.
 
The other symbols in eqs. (8)--(10) describe the mass, energy and
momentum source terms due to radiative and mechanical feedback, and
are described in Sects. 3.3 and 3.4 below.  In particular, the terms
$\rhodotw$ and $\vvw$ are given in
eqs. (25)--(28).  $\gradient p_\mathrm{rad}=(\gradient
p_\mathrm{rad})_\mathrm{photo} + (\gradient
p_\mathrm{rad})_\mathrm{es}$ is the total radiative pressure gradient
for the radiation coming from the accreting MBH.  Most of the input
physics concerning feedback and SF is the same as in Ciotti et
al. (2010), with a few modifications described in detail below.

\subsection{Star formation}

Star formation, and the consequent SNII production, are treated as in Novak et al. (2011)
and Negri et al. (2015). SF is implemented by subtracting gas from the grid, and this mass
sink  corresponds to a SF rate per unit volume given by:
\begin{equation}
  \dot{\rho}_\mathrm{SF} = \frac{\eta_{\rm SF} \rho }{\tau_{\rm SF}} \, ,
\end{equation}
where $\rho$ is the local gas density, $\eta_{\rm SF}$ is the SF efficiency, for which
we adopt the value of 0.1, and 
\begin{equation}
  \tau_{\rm SF} = \max(\tau_{\rm cool}, \tau_{\rm dyn}),\quad
\end{equation}
where
\begin{equation}
  \tau_{\rm cool} = {E\over C}, \quad
  \tau_{\rm dyn} = \min(\tau_{\rm jeans}, \tau_{\rm rot}),\quad
\end{equation}
with
\begin{equation}
  \tau_{\rm jeans} = \sqrt{\frac{3\pi}{32 G \rho}}, \quad   \tau_{\rm rot} = \frac{2\pi r}{v_c(r)}.
\end{equation}
Here $G$ is the Newtonian gravitational constant, $r$ is the distance from the galaxy center, 
$\tau_{\rm rot} $ is an estimate of the radial epicyclic period, and
$v_c(R)$ is the galaxy circular velocity in the equatorial plane. The energy and momentum 
sinks associated with star formation are then:
\begin{equation}
 \dot{E}_\mathrm{SF} = \dfrac{\eta_\mathrm{SF} E}{\tau_\mathrm{SF}},  \qquad \dot{\boldsymbol{m}}_\mathrm{SF} = \dfrac{\eta_\mathrm{SF}\boldsymbol{m}}
{\tau_\mathrm{SF}} = {\dot{\rho}}_\mathrm{SF}\uv,
\end{equation}
where $E$ and $\boldsymbol{m}$ are the internal energy and momentum density of
the ISM.

SF removes mass, momentum and energy from the grid, but also injects
new mass and energy from SNII explosions.  For each SF episode,
assuming that the new stars form with a Salpeter IMF, the mass
returned in SNII events is 20\% of the new star mass in that episode;
the SNII mass source term $\dot \rho_{\rm II}$ at each time $t$ comes
from considering that a given SF episode generates SNII's that inject
mass (at a rate exponentially declining on a timescale $\tau_{\rm II}
= 2\times 10^7$ yr), and that during the evolution of that episode
other episodes may take place, forming other SNII's that in turn eject
mass into the ISM. The same considerations are taken into account to
compute the SNII energy injection rate ${\dot E}_{\mathrm{II}}$ (see
Negri et al. 2015 for more details on how $\dot \rho_{\rm II}$ and
${\dot E}_{\mathrm{II}}$ are computed).

\subsection{Mechanical Feedback}
Following the general, self-consistent treatment in Ostriker et al. (2010), the basic quantities involved 
in the implementation of mechanical feedback are written as:
\begin{equation}
 \Mdotbh  = \dfrac{\Mdotin}{1+\eta}, 
  \label{eq:mdot-bh}
\end{equation}
\begin{equation}
\Mdotout = \eta \Mdotbh,
 \label{eq:mdot-w}
\end{equation}
\begin{equation}
 \Lw = \epsw \Mdotbh c^2,
   \label{eq:edot}
\end{equation}
\begin{equation}
 \pdotw = \Mdotout  \vw ,
 \label{eq:pdot}
\end{equation}
where $\Mdotbh$ is the mass accretion rate on the MBH, $\Mdotout$ is
the mass outflow rate in the conical wind, $\Mdotin$ is the mass
inflow rate at the first active radial grid, $\epsw $ is the
efficiency of generating mechanical energy with an AGN wind, $\vw
=\Vert\vvw\Vert$ is the modulus of the AGN wind
velocity, and $\eta \equiv 2\epsw c^2 / \vw^2$.  With the adoption of
this simplified scheme, the sub-grid physics near the MBH (described
in Ciotti \& Ostriker 2012) is not activated here. Then, in the
present simulations, all gas, that flows in, eventually either flows
onto the MBH or back into the simulation grid as a conical wind,
without passing through a circumnuclear starforming disk.

Equations (\ref{eq:mdot-bh})–(\ref{eq:pdot}) above are adopted by most
authors to treat AGN feedback as a process comprising both infall and
outflow; however, typically $\eta = 0$ is adopted, implicitly assuming
$\vw\rightarrow \infty$, so that $\Mdotout$ and
$\pdotw$ are neglected, and $\Lw$ and $\Mdotbh$
may be overestimated.  For example, if we adopt $\epsw =5 \times
10^{-3}$ as many authors [e.g., Springel et al. (2005), Johansson et
al. (2009), McCarthy et al. (2010)], and $\vw = 10^4~\kms$
(e.g., Moe et al. 2009), then $\eta =9$, and all the normally
neglected effects may in fact be dominant: the bulk of the inflowing
mass ($\Mdotin$) may be ejected in a broad-line disk wind,
and the effects of the mass ($\Mdotout$) and momentum
($\pdotw$) input deposited in the ambient gas may dominate
over the energy input ($\Lw$), which may be largely radiated
away. Thus, it is important to consider consistently the effects of
including mass, energy, and momentum conservation when $\eta > 0$; our
treatment is consistentwith observations of BAL quasars and the
radiative mechanisms that drive them (Ciotti et al. 2010).

In general, $\eta$ is fixed for given $\epsw$ and $\vw$.  The wind
efficiency, $\epsw$, is not known very well, neither from observations
nor from detailed simulations. The best estimates might be in the
range $10^{-3}> \epsw > 3 \times 10^{-4}$ (Proga et al. 2000; Proga \&
Kallman 2004; Krongold et al. 2007; Kurosawa et al. 2009). Here we
consider models where $\epsw$ and $\vw$ are allowed to depend on the
mass accretion rate, and are described by the following laws (see
Novak et al. 2011 for a discussion on these assumptions):
\begin{equation}
\epsw =  \dfrac{\epswz \Aw\mdot}{1+\Aw\mdot}, \qquad 
\vw = \dfrac{\vwz \Aw\mdot}{1+\Aw\mdot}, 
\end{equation}
where $\Aw =1000$, $\mdot$ is the dimensionless mass accretion rate,
i.e., normalized to the Eddington mass accretion rate (see below), and
the two functions $\epsw $ and $\vw$ saturate to $\epswz$ and $\vwz$,
respectively, for large accretion rates. We choose the constant values
$\epswz =10^{-4}$ and $\vwz =10^{4}$~km/s (as observed for the outflow
velocity in UV absorption lines of BAL AGNs; e.g. Reichard et
al. 2003, Gibson et al. 2009; in ionized emission of high-redshift
quasars, e.g. Liu et al. 2013, Zakamska et al. 2016; in nearby Seyfert
galaxies, e.g. Fischer et al. 2013; and in molecular outflows, Tombesi
et al. 2015, Feruglio et al. 2015).  Independently of the mechanical
feedback model, the radiative luminosity of the AGN is given by:
\begin{equation}
  \Lbh = \epsem\Mdotbh c^2  ,
  \label{eq:lum}
\end{equation}
where the electromagnetic efficiency $\epsem$ is given by the
advection dominated accretion flow inspired formula (Narayan \& Yi
1995), to reproduce even the very low nuclear luminosities typically
observed for local MBHs (e.g. Pellegrini 2005):
\begin{equation}
\epsem = \frac{\epsemz\Aem\mdot}{1 +\Aem\mdot} ,
  \label{eq:epsilon-em}
\end{equation}
and $\Aem =100$ and $\epsemz =0.125$.  The dimensionless mass accretion rate is
\begin{equation}
\mdot\equiv \frac{\Mdotbh}{\Mdotedd} = 
  \frac{\epsemz\Mdotbh c^2}{\Ledd} \, ,
\end{equation}
where $\Ledd$ is the Eddington luminosity. With the settings above, $\eta$ has a minimum value of 0.18 at the highest
accretion rates, and increases for decreasing $\mdot$ (for example,
$\eta=0.36$ for $\mdot =10^{-3}$).

Our previous 1D simulations used a prescription based on pressure
balance between the outgoing wind and the ambient gas, to compute how
the mass, energy, and momentum are radially distributed; the present
2D simulations instead inject the desired mass, energy, and momentum
into the innermost radial cells, and self-consistently compute the
radial transport of these quantities.  In the 2D simulations, the {\it
  total} mass, energy and momentum injected into the ISM by the AGN
wind are calculated as described by eqs. (18)-(20), for a given MBH
accretion rate $\Mdotbh$; then, the specification of the angular
dependence of the properties of the AGN conical wind
($\rhodotw$, $\dot{{\bf m}}_\mathrm{w}$ and $\dot
E_\mathrm{w}$) is required.  For this dependence we adopt:
\begin{equation}
\rhodotw = \Mdotout {\delta(r-r_0) \over r^2 {\rm sin}\theta} f(\theta),
\end{equation}
where $\delta $ is the Dirac delta-function, $r_0$ is the first gridpoint, $\theta$ is the angle from the $z$-axis, measured
clockwise, and:
\begin{equation}
f(\theta)= { (n+1) \, {\rm sin}\theta \, |\cos\theta|^n  \over 4\pi}. \label{eq:angular_dist}
\end{equation}
The anisotropic momentum source in eq. (9) is given by:
\begin{equation}
\dot{{\bf m}}_\mathrm{w} = \rhodotw\vw {\bf e}_r =\pdotw 
{\delta(r-r_0) \over r^2 {\rm sin}\theta} f(\theta) {\bf e}_r
\end{equation}
where the second equality follows from using eqs. (21) and (26). Finally, the energy injection due to the AGN wind,
from eqs. (20) and (26),  is:
\begin{equation}
\dot E_\mathrm{w} = {1\over 2} \rhodotw \vw^2 = \Lw 
{\delta(r-r_0) \over r^2 {\rm sin}\theta} f(\theta).
\end{equation}
Note that $\vw$ is independent of $\theta$, and the anisotropy in the momentum and energy injection 
is due to the anisotropic mass injection [eq. (26)].
Following Novak et al. (2011), we adopt $n=2$,
so that the half-opening angle enclosing half of the energy is $\sim$
$45^\circ$.  In terms of solid angle, this means that the wind is
visible from $\sim 1/4$ of the available viewing angles.  This
fraction is in agreement with observations of the fraction of obscured
and unobscured AGNs under the assumption that the two populations are
made up of a single population of objects that differ only in viewing
angle (e.g.  Liu et al. 2015; Bae \& Woo 2016).

\subsection{Radiative Feedback: heating and cooling}

Radiative heating and cooling are computed by using the formulae in Sazonov
et al. (2005), which describe the net heating/cooling rate per unit volume of a
plasma in photoionization equilibrium with a radiation field characterized
by the average quasar Spectral Energy Distribution (Sazonov et al. 2005, 2008), whose associated spectral 
temperature is $\Tc \simeq 2$ keV. In particular,
Compton heating and cooling, bremsstrahlung losses, line and continuum heating
and cooling, are taken into account.
The net gas energy change rate per unit volume for $T \gtrsim 10^4$~K
is given by:
\begin{equation}
 H-C \equiv n^2 (S_1 + S_2 + S_3),
\label{dotE}
\end{equation}
where $n$ is the Hydrogen number density, and positive and
negative terms are grouped together in the heating ($H$) and cooling
($C$) functions (all quantities are expressed in cgs system). The bremsstrahlung losses are given by
\begin{equation}
S_1 = -3.8\times 10^{-27}\sqrt{T}.
\end{equation}
The Compton heating and cooling is given by
\begin{equation}
S_2 = 4.1\times 10^{-35} (\Tc -T)\,\xi ,
\label{eqS2}
\end{equation}
and $\xi$ is the ionization parameter. The
sum of photoionization heating, line and recombination continuum cooling is
\begin{equation} 
S_3 = 10^{-23}\dfrac{a + b\, (\xi/\xi_0)^c}{ 1 + (\xi/\xi_0)^c}\dfrac{Z}{Z_\sun},
\label{eqS3}
\end{equation}
where the almost perfect linear dependence on metallicity is explicit, and
\begin{equation}
a=-\dfrac{18}{  e^{25 (\log T -4.35)^2}} 
    -\dfrac{80}{  e^{5.5(\log T -5.2)^2}}
    -\dfrac{17}{  e^{3.6(\log T -6.5)^2}},
\end{equation}
\begin{equation}
b=1.7\times 10^4\;T^{-0.7},  \qquad 
c=1.1-\dfrac{1.1}{ e^{T/1.8 \times 10^5}}+\dfrac{4\times 10^{15}}{T^4}, 
\end{equation} 

\begin{equation}
\xi_0 = \left( \dfrac{1.5}{ T^{0.5}}+\dfrac{1.5\times 10^{12}}{ T^{2.5} }\right)^{-1} +
      \dfrac{4\times 10^{10}}{ T^2}
          \left[1 + \dfrac{80}{ e^{(T-10^4)/1.5\times 10^3}}\right].
\end{equation}
Gas temperatures are bounded from below by the adopted atomic
cooling curve, that has an exponential cutoff below $10^{4}$ K.

In this work we do not consider the effects of dust, i.e., the radiation momentum associated with dust absorption (cf.~Debuhr
et~al. 2011; Bieri et al. 2016); however, the effect of dust has been shown not to alter significantly the gas evolution (Hensley et al. 2014).
Moreover, since the effects of dust absorption and reprocessing are not considered here, 
the radiation pressure due to the reprocessing of light emitted by stars formed during the evolution is also omitted.
For the radiation pressure we consider electron scattering and the force exerted by absorption of AGN photons by atomic lines 
(see eq. 55 in Ciotti \& Ostriker 2007). The integration scheme used here for the radiation transport is not the full scheme of
Novak et al. (2012, their Sect. 3), but the simplified version in their Appendix B, in order to speed up the simulations.
For example, in the case of spherical symmetry, the equation to be integrated would be:
\begin{equation}
\dfrac{d\lbhefphot (r)}{ dr}=-4\pi r^2 H,
\end{equation}
where $\lbhefphot (r)$ is the effective accretion luminosity at $r$,
and the equation is solved with central boundary condition $\lbhefphot (r=0)=\Lbh$ given by
equation~(\ref{eq:lum}).
The force per unit mass due to photoionization+Compton opacity can be expressed as 
\begin{equation}
(\gradient p_\mathrm{rad})_\mathrm{photo} = - \dfrac{\rho \kappa_\mathrm{photo}}{c} \dfrac{\lbhefphot (r)}{4\pi r^2} {\bf e}_r,
\end{equation}
where 
\begin{equation}
\begin{split}
\kappa_\mathrm{photo}=-\dfrac{1}{\rho(r)\lbhefphot (r)} \dfrac{d\lbhefphot (r)}{ dr}= \\
      =\dfrac{4\pi r^2 H(r)}{\rho(r)\lbhefphot (r)}.
\end{split}
\end{equation}

The radiation pressure due to electron scattering is 
\begin{equation}
(\gradient p_\mathrm{rad})_\mathrm{es} = - \dfrac{\rho \kappa_\mathrm{es}}{c} \dfrac{\Lbh}{4\pi r^2}{\bf e}_r,
\end{equation}
where $\kappa_\mathrm{es} = 0.35~\mathrm{cm^2~g^{-1}}$.

Finally, equations~(\ref{eqS2}) and (\ref{eqS3}) depend on the ionization parameter, that is given by:
\begin{equation}
\xi \equiv \dfrac{\lbhefphot (r)}{ n(r) r^2}.
\end{equation}

\section{The simulations}

We employed our modified version of the parallel ZEUS code (Hayes et
al. 2006), in a 2D axisymmetric configuration, with a radially
logarithmic grid in spherical coordinates $(r,~\theta)$ of $128 \times
32$ meshpoints, spanning from 2.5~pc to 250~kpc.  Reflecting boundary
conditions were set along the $z$-axis, while at the outer edge of the
computational domain the fluid is free to flow out.

Negri et al.~(2014a,b) showed that in a galaxy with substantial
ordered rotation (without AGN feedback) the gaseous halo is almost
co-rotating with the stars, and angular momentum conservation leads to
the formation of a cold, rotationally supported, star forming
equatorial disk of kpc-scale (Negri et al. 2015), thus preventing any
substantial accretion on the central MBH. These massive disks are
expected to be gravitationally unstable, fragment, and consequently
transport material to the galaxy center (Bertin \& Lodato 2001;
Hopkins \& Quataert 2011).  At the present stage we do not account for
these processes, and we restrict to the low-rotation case.  However,
in axysimmetric systems some rotation is numerically needed to prevent
gas from unphysically sticking on the $z$-axis.  Different recipes
have been figured out to solve this problem (Novak et al. 2011; Li,
Ostriker \& Sunyaev 2013); here, we take advantage of the Jeans solver
that allows us to tune the Satoh $k$-parameter (Sect. 2.2). In detail,
we determine $k(r,\theta)$ in order to have negligible but non-zero
ordered rotation over the main galaxy body, with angular momentum of
the stars never exceeding $J_\mathrm{0}$, the specific angular
momentum of the circular orbit at the first radial gridpoint, in the
gravitational field of the MBH. In this way the gas is allowed to
enter the first gridpoint, and so be accreted on the MBH.  In
practice, we consider a low rotation regime, where the centrifugal
barrier prevents gas from sticking onto the $z$-axis, but allows for
accretion down to the innermost radial gridpoint; such a rotation
field is built by defining the Satoh parameter as follows:
\begin{equation}
 k(r,\theta) \equiv  {\eta_{rot} J_{\mathrm{0}} \over \sqrt{ J_{\mathrm{0}}^2 + R^2 v_{\mathrm{IS}}^2} }; \qquad v_{\mathrm{IS}}^2 \equiv 
\overline{\vphi^2}-\sigma^2,
\end{equation}
where $R=r\,{\rm sin}\theta$ is the distance from the $z$-axis, 
$v_{\mathrm{IS}} (r,\theta)$ is the ordered stellar velocity if the galaxy model were an isotropic rotator, and we fix $\eta_{rot}=0.9$. 

We computed various X-ray properties of the gas flows, as 
the X-ray luminosity in the 0.3-8~keV band, and the
X-ray-emission-weighted temperature, respectively defined as:
\begin{equation}
 \Lx = \int \varepsilon_\mathrm{X} dV, \qquad \Tx = \dfrac{1}{\Lx}\int T\varepsilon_\mathrm{X} dV ,
\end{equation}
where $\varepsilon_\mathrm{X}$ is the thermal emissivity in the 0.3-8~keV band of a hot, collisionally ionized plasma 
(see Negri et al. 2014a for more details), and the 
integrals are computed over the volume of interest. We also calculated the X-ray surface brightness maps and 
maps of projected temperatures (as detailed in Pellegrini et al. 2012).
We also calculated the following general properties of the models: the duty cycle ($\mathcal{D}$), defined as the percentage of time 
spent at a $l \equiv \Lbh /\Ledd >0.05$; the time at which half of the MBH radiation energy has been
emitted ($t_L$); the time at which half of the new stellar mass has been created ($t_M$); 
and the half-mass radius of the new stars formed until the end of the simulation ($r_M$).

We assume that each galaxy at the beginning of the simulation is
2~Gyr old, and is depleted of gas due to the intense high star formation occurring 
during the initial stages of its evolution. The simulations follow the galaxy evolution for the subsequent
11 Gyr, thus ending at a galaxy age of 13 Gyr.

\section{Results}

We present here the main results for the whole set of eight galaxy
models in Tab. 1. For each galaxy model we ran three simulations: one without
any feedback from the MBH (we refer to these models as to NOF models), one with
mechanical feedback only (MF models), and one where the feedback is 
radiative plus mechanical, due to an AGN conical wind (full-feedback FF models). The most interesting
quantities at the end of the twenty-four simulations are listed in
Tab. 2. Rather than presenting the specific evolution of the gas flow in each one of
the various models, in the following we focus on a representative model, and then
we discuss the overall results across the whole set, as a function of
galaxy mass, shape and type of feedback. We focus on the results concerning the
the hot gas (Sect. 5.1),
the MBH growth (Sect. 5.2), the newly formed stars (Sect. 5.3), and the
duty cycle (Sect. 5.4).

\subsection{Gas evolution}

During quiescent phases the flow is characterized by a central
inflowing region, and an external outflowing one, as found in our
previous studies (e.g. Ciotti et al. 2010 for spherical models; Negri
et al. 2014b for 2D simulations without AGN feedback).  The main
effect of varying the galaxy mass is that the size of the central
inflow is larger for more massive galaxies, with consequences for the
gas content, the average gas density and cooling time, the mass
accretion rate and the related quantities (as discussed below). At
fixed galaxy mass, a change in the galaxy shape produces a lower $\Lx$
for flatter morphologies, because the outflowing region becomes
larger, as obtained for 2D models without feedback (Negri et
al. 2014b).

\begin{figure*}
\vskip -2truecm
\hskip -1truecm
\includegraphics[scale=0.9]{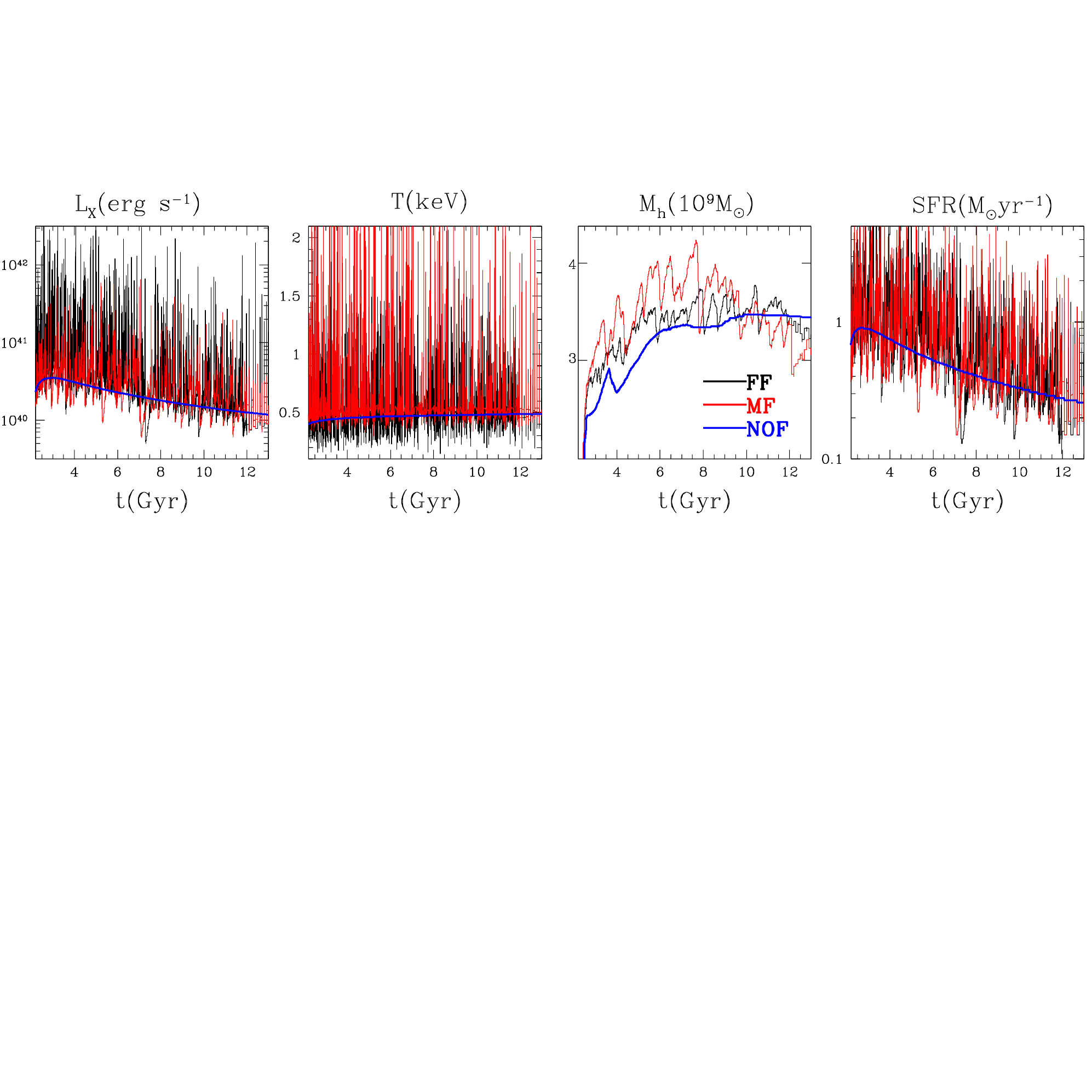}   
\vskip -12 truecm
\hskip -1truecm
\includegraphics[scale=0.9]{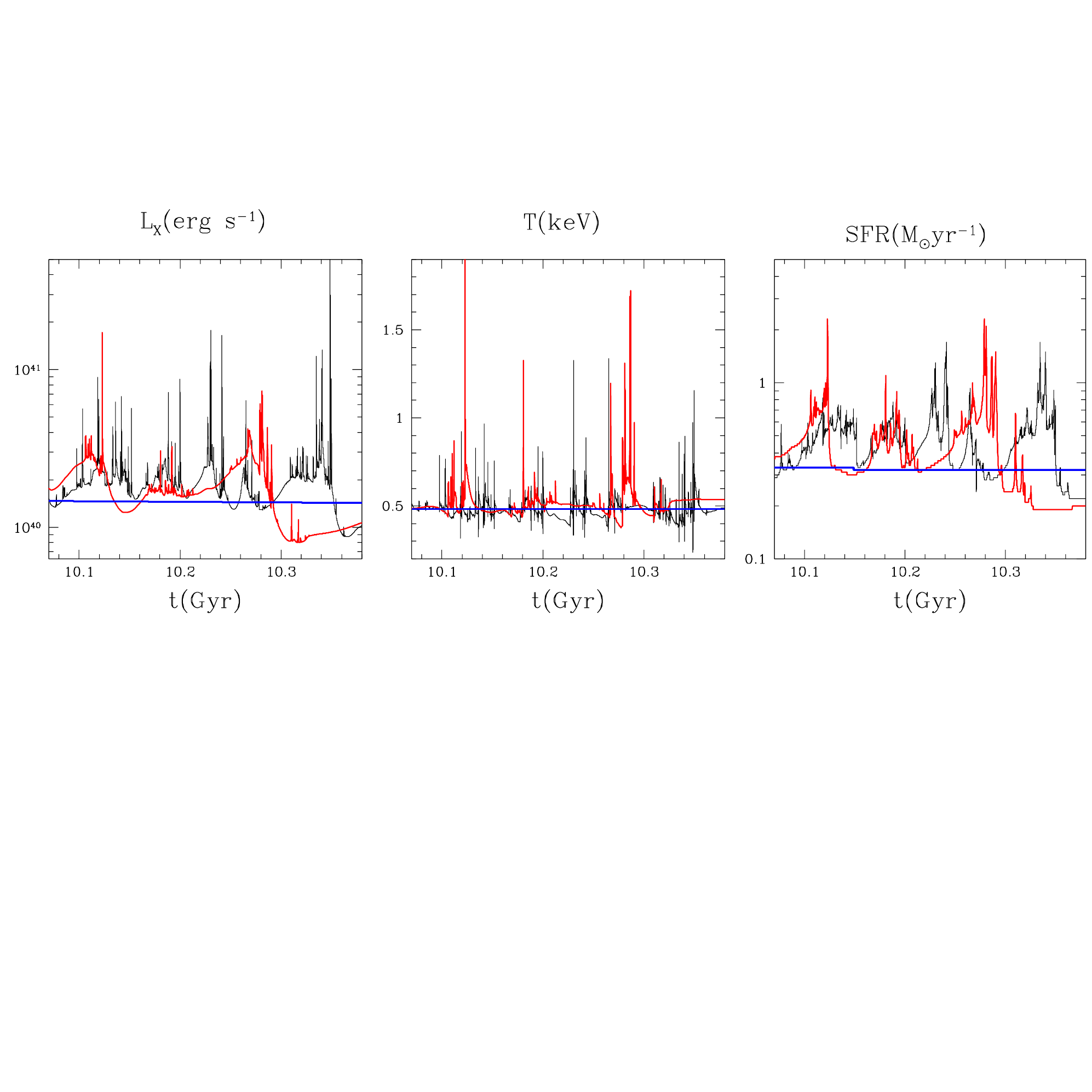}   
\vskip -7.5truecm
\caption{Time evolution of the main gas flow properties, for the E4$^{210}$ galaxy model (see Tabs. 1 and 2).
From left to right, plotted quantities are the X-ray luminosity and emission weighted temperature, in the 03--8 keV band, and
within 5$\re$, the mass of the hot gas within the whole numerical grid, and the star formation rate.
Blue, red and black lines  refer to no feedback (NOF), mechanical feedback only (MF) and full feedback (FF) models.
In the bottom panels, a zoom in time is made between 10 and 10.4 Gyr.}
\label{f1} 
\end{figure*}

We start considering a representative E4 model of average mass
(E4$^{210}$ in Tab. 1). This choice is motivated by the fact that the
morphological E4 type of ETGs is observed to be more common than the
E7 one, and also that our galaxy models (that are essentially
non-rotating) are more realistic for E4 galaxies than for E7 ones.
Figure~\ref{f1} shows the time evolution of some quantities describing
the hot gas phase in the E4$^{210}$ galaxy: $\Lx$ and $\Tx$ calculated
within 5$\re$, the hot gas mass within the whole grid, and the SFR.
The NOF model shows a smooth evolution in $\Lx$, $\Tx$, hot gas mass
$M_{h}$ (the mass with $T>10^6$ K), and SFR; in particular, in each
panel of Fig.~\ref{f1}, the NOF curves represent a sort of average
lower envelope for the largely fluctuating behavior typical of
feedback models. The present-epoch values of $\Lx$, $\Tx$, $M_{h}$ and
SFR are not significantly different between the NOF, MF and FF cases;
this is true for the whole set of models, at all galaxy masses and
shapes (see also Tab. 2).  Therefore, we can conclude that the global
properties of the hot gas are not significantly affected by AGN
feedback.  However, we note that after each outburst $\Lx$ of feedback
models, when {\it calculated within} $(1-2)\re$, shows drops that
reach one order of magnitude below $\Lx$ of NOF models, i.e. that are
much larger than those in Fig. 1. The drops are the natural
consequence of the clearing of the gas in the central regions.

Figure 2 shows the $\Lx$ vs. $\Tx$ relation for all models, at the
present epoch, in comparison with the observed $\Lx$ and $\Tx$ values
recently measured for the gas only, using $Chandra$ data (Kim \&
Fabbiano 2015, Goulding et al. 2016).  The figure shows how the NOF,
MF and FF models occupy similar regions in the plot, and also a
remarkable agreement between models and real ETGs. In particular, the
large spread in $\Lx$ at low luminosities is well reproduced, as due
to mostly outflowing ETGs at lower galaxy masses.  Also well
reproduced is the observed average trend for $\Lx \ga 10^{40}$ erg
s$^{-1}$ (note that most of the highest $\Lx$ observed are due to
central ETGs in groups, Goulding et al. 2016).

The time evolution of $\Lx$, $\Tx$, $M_{h}$ and SFR is similarly rich
in large fluctuations in models more massive than the E4$^{210}$ one,
while it generally becomes smooth and slowly declining in less massive
models, where a global outflow is established within the present
epoch. Massive ETGs, then, may appear to spend much of their lifetime
at very high $\Lx$, $\Tx$ and SFR values (Fig. 1). This aspect can be
checked quantitatively by computing the gas ``duty-cycle'', as the
fraction of time, during a chosen range of time, spent by a certain
gas quantity at a level above a chosen average value for the NOF case.
These fractions turn out to be very low. For example, for the gas
$\Lx$, one can compute the fraction of time spent above $\Lx=3\times
10^{41}$ erg s$^{-1}$, that represents a sort of upper value for the
observed gas emission of normal ETGs in the local
universe\footnote{Note that this is conservative assumption; for
  example, central group or cluster ETGs can show much larger $\Lx$
  values.} (Kim \& Fabbiano 2015), over the past 3 or 5 Gyr. For both
choices of lookback time, the E4$^{250}$ and E7$^{250}$ FF models
spend $\sim 1$\% of the time with $\Lx>3\times 10^{41}$ erg s$^{-1}$,
and lower mass models spend $<1$\% of their time above it (the very
massive E4$^{300}$ and E7$^{300}$ models have $\Lx$ precisely of the
order of $2-3\times 10^{41}$ erg s$^{-1}$ in the past few Gyr; see
Tab. 2).
Of course, the fraction of time during which 
disturbances in the gas remain can be larger than this, as already discussed for spherical models by Pellegrini et al. (2012).


The large fluctuations in the gas properties are produced by nuclear
outbursts. For illustrative purposes, we have selected one
representative outburst, well isolated in time, taking place at 6.85
Gyr for the E4$^{180}$ FF model; in Figs.~\ref{dens}$-$\ref{sfrd} we
show the evolution of the main hydrodynamical quantities during this
outburst.  The leftmost panels in these figures show the quiescent
time closest to the outburst, when the gas properties still have a
smooth distribution. In the middle panels, corresponding to subsequent
times, the gas reaches its peak in emission, and cold and dense
fingers are approaching the galactic center (Figs.~\ref{dens}
and~\ref{temp}), where outflow and inflow regions coexist
(Fig.~\ref{vr}); these cold filaments are mixed with hot and low
density regions already created by the developing outburst. The
outburst is then fading; there is still some outflowing material from
the nucleus, and some hot, low density outflowing material at a radius
of $\la 1$ kpc.  The rightmost panels show again a quiescent state, at
a time of 6.95 Gyr, i.e., $10^8$ yr after the start of the burst; the
galaxy still has a lower density and higher temperature gas than right
before the outburst. Close to the nucleus one can now notice the
heating effect of the fading AGN wind, as a slightly hotter, lower
density bi-conical region (Figs. 3 and 4); the wind itself is
outflowing in a small hourglass region above and below the nucleus
(Fig. 5), and imparts a tangential velocity to the gas surrounding it
(Fig. 6).

Figure 7 shows the SFR during the outburst: SF is very active in the
cold filaments close to the center; at the end of the burst, one can
see a lower SFR region close to the nucleus, due to the AGN wind. This
feature nicely agrees with an observed spatial anti-correlation
between H$\alpha$ emission, a tracer of SF, and line emission from
powerful outflows in quasars (Cano-D\'iaz et al. 2012, Cresci et
al. 2015, Carniani et al. 2015); this observation is considered an
important evidence for negative feedback, however we find that overall
the feedback effect on SF is positive (Sect. 5.3).  Figure~\ref{brill}
shows finally the X-ray surface brightness and projected temperature
maps weighted with the emission over 0.3--8 keV (Sect. 4),
corresponding to the two central times during the outburst of
Figs. 3--7 (t=6.85 and t=6.86 Gyr). A very bright central region of
$\sim 1$ hundred of pc radius is apparent in the brightness map; sharp
and very hot arcs can be seen in the temperature map. Both features
are very transient, but should be detectable with $Chandra$'s high
angular resolution in nearby galaxies; similar features were found in
1D simulations discussed in Pellegrini et al. (2012).

We note that the bursting activity is almost continuous and lasts for
the whole lifetime of massive models, at variance with the activity
shown by 1D hydrodynamical simulations, that was characterized by
peaks well isolated in time, and decreasing in frequency at later
times (Ciotti \& Ostriker 2012).  This major difference is due to
obvious geometrical reasons: in spherical symmetry the cold shell,
produced by the snowplow phase of AGN-launched shock waves, cannot
fragment; as a result, the shock waves are very efficient in clearing
the inner regions of the galaxy from the gas. Instead, 2D
hydrodynamics allows for Rayleigh-Taylor instabilities of the cold
shell, that breaks ``permitting cold fingers of material to accrete
onto the BH'' (Ciotti \& Ostriker 2001), while hot gas is escaping
from the center at the same time. This behavior was also found in
spherical models simulated with the 2D code (Novak et al. 2012); the
possibility of multiphase and cold gas accretion on to a MBH has been
intensively investigated recently with a variety of numerical
simulations (e.g. Pizzolato \& Soker 2010; Barai, Proga \& Nagamine
2012; Nayakshin \& Zubovas 2012; Gaspari, Ruszkowski \& Oh 2013).

Finally, during a burst episode we find no major differences between
the FF and MF cases, in the behavior of the hydrodynamical quantities
on the galactic scale; the time evolution of central or nuclear
properties, instead, as the gas emission during its peaks, or the
nuclear emission, is more structured in the FF than in the MF case.
This results in a longer duration, on average, of the outbursts in FF
models, as shown in the lower panels of Fig. 1, that gives an example
of this difference in time evolution, with a zoom in between 10 and
10.4 Gyr. This property, already shown by spherical models (Ciotti,
Ostriker \& Proga 2009, 2010), has interesting consequences for the
cumulative SF of MF and FF models (Sect. 5.3).

\subsection{The black hole mass growth}

Figure~\ref{mbh} shows the adopted initial $\Mbhz -\Mstar$
relation (Sect. 2.2), the $\Mbh-\Mstar$ relation derived from
dynamical studies of well observed local ETGs (McConnell \& Ma 2013),
and the $\Mbh$ values at the end of the simulations (when the MBH mass
has grown, while $\Mstar$ remains constant); different symbols
distinguish the galaxy shapes and types of feedback. The first thing
to notice is that models with feedback produce a final $\Mbh -\Mstar$
relation that agrees well with the observed one, within the
uncertainties.  This agreement is not fulfilled by NOF models that end
with overwhelmingly large $\Mbh$ masses (not shown in the figure, but
see Tab. 2), thus demonstrating once again how feedback is needed to
keep the MBH masses reasonable, even after ETGs have become ``red and
dead''.

MBH masses measured locally and those of feedback models are
consistent at all galactic $\Mstar$ and shapes; in fact, the
differences between the final $\Mbh$ values at fixed $\Mstar$, due to
shape and type of feedback, are minor. This shows that the effect on
the MBH mass growth of mechanical feedback is very important (even
though this does not imply that the radiative feedback alone has minor
effects). Looking deeper into Fig.~\ref{mbh}, however, one can notice
some trends at fixed $\Mstar$. E7 models show lower $\Mbh$ than E4
ones, a result due to the lower gas binding energy of flatter galaxies
(Ciotti \& Pellegrini 1996, Posacki et al. 2013), with the consequent
larger outflow region with respect to central inflowing one, and
larger effectiveness of AGN feedback, leading to a lower accreted
mass. For the same galaxy shape, FF models end with lower $\Mbh$ than
MF ones (Fig.~\ref{mbh}), a result of their larger capability of
preventing gas from accreting.  As a future investigation, it would be
interesting to study whether the final MBH masses keep close to those
measured locally even when starting from different (lower) values for
$\Mbhz$, and more in general what is the effect of the
initial (albeit at $z\sim 2$) value of $\Mbhz$ on the flow
evolution and on the final $\Mbh$. However, simulations with smaller
$\Mbhz$ are more time-consuming, if, as done in all our
studies, one wants to resolve the Bondi radius, and the even smaller
radius at which the thermal energy associated with the AGN spectral
temperature equals the particle's gravitational energy in the MBH
potential (Ciotti \& Ostriker 2001).

The increase in $\Mbh$, for feedback models, is different for
the different galaxy masses: it ranges from a figure of the order of
$\sim 10$\% for the lowest mass models (E$^{180}$), to roughly $\sim
100$\% (the MBH mass $\sim$doubles) for the E$^{210}$ models, to
roughly triplicate for the E$^{250}$ models, to almost quadruplicate
for the very massive E$^{300}$ galaxies (see col. 3 in Tab. 2, and
Fig.~\ref{mbh}, right panel). Therefore, MBHs in more massive ETGs not
only start larger and are still larger at the end of the simulations,
but are also able to grow more in percentage, with respect to their
initial mass.  This is because more massive MBHs have larger Eddington
luminosities, and, most importantly, they reside in ETGs where the gas
is more bound, per unit mass; then, more massive ETGs have larger
accreting gas mass per unit stellar mass, or per unit MBH mass.  At
fixed $\Mstar$, the right panel of Fig.~\ref{mbh} shows the same
trends shown in the left panel: the MBH mass increases more, even with
respect to the initial ${\rm M_{BH,0}}$, in E4 models than in E7 ones,
and in MF models than in FF ones. Such a mass increase may appear
large, but in fact is remarkably small: in absence of AGN feedback and
SNIa's assisted galactic winds, the mass increase of $\Mbhz$
would be larger by up to two orders of magnitude (Tab. 2).  It is a
valuable property of the present models that the average increase in
$\Mbhz$ since $z\sim 2$ is just of a factor of few ($2-4$), as
found for the average growth history of MBHs with given starting mass
(e.g. Marconi et al. 2004).

\subsection{The newly formed stars and the circumgalactic medium}

Figure~\ref{newst} (left panel) shows the mass in newly formed stars,
at the end of the simulations, for models with feedback.  $\DMstar$
ranges from $\simeq 6\times 10^8\Msun$ to $\simeq 3\times
10^{10}\Msun$, from the low to the high galaxy masses.  The
lowest mass models show significantly lower $\DMstar$
because during their evolution the gas is mostly outflowing, they
experience little accreting mass and little possibility of SF.  At
fixed $\Mstar$, rounder models produce slightly larger $\DMstar$ than
flatter ones, due to their larger capability to retain the gas, that
can eventually be converted into stars.  MF and FF models produce
similar $\DMstar$, at any fixed $\Mstar$ and galaxy shape, with larger
$\DMstar$ in the FF case. The larger efficiency of FF models to
produce SF is a consequence of the richer structure and longer
duration of each of their nuclear outburst episodes, that prevent
prompt accretion, provide the gas more time for its cooling, and lead
to a larger capability to keep cold and dense gas far from the nucleus
(see Sect. 5.1, and Fig. 1, where an FF outburst shows more
fluctuations in $\Lx$, and a larger duration of the feedback action on
the gas, than a MF outburst).  Note that the final $\DMstar$, at fixed
$\Mstar$ and galaxy shape, increases from NOF to MF to FF models
(Tab. 2, except for the two NOF E$^{180}$ models), which proves
definitely that feedback has the overall effect of making SF more
efficient, i.e., it has a {\it positive} action with respect to SF.
Also, the behavior of the $\DMstar$ growth is opposite to that of the
MBH growth, with respect to feedback: $\DMstar$ increases from NOF to
FF models, while the opposite is true for $\DMbh$ (cfr. Figs. 9 and
10).  Both $\DMstar$ and $\DMbh$, instead, are larger for rounder
models.

It is interesting to compare the relative increase of $\DMstar$ and
$\DMbh$ (values in Tab. 2).  In NOF models, $\DMbh$ and $\DMstar$ are
not largely different, and the MBH growth is much favored over the SF,
with respect to feedback models.  For feedback models, $\DMbh$ ranges
from $\simeq 6\times 10^6\Msun$ to $\simeq 2\times
10^{9}\Msun$, thus $\DMstar$ is always much larger than $\DMbh$,
at all galaxy masses.  There is a trend, though: $\DMbh/\Mbhz$ (Fig.~\ref{mbh}) increases clearly with $\Mstar$,
while $\DMstar/\Mstar$ keeps quite flat or even decreases
(Fig.~\ref{newst}, right panel). Thus, $\DMbh/\DMstar$ increases with
$\Mstar$, and there is relatively more MBH growth than SF in more
massive galaxies.  The reason for this trend is that the mass flowing
to the central region, where SF takes place (see below), is converted
in new stars more efficiently than in MBH mass at low galaxy masses,
while the opposite works in the most massive galaxies. In more massive
ETGs, in fact, the feedback can displace the gas out to distances from
the MBH that are relatively lower than in less massive ones,
consequently the gas has less time to fall back to the center, while
giving origin to SF (as revealed by an inspection of the evolution of
the hydrodynamical properties of the flow).  One final remark here
concerns the Magorrian relation $\Mbh - \Mstar$: as the stellar
mass does not increase significantly after $z\sim 2$
(Fig.~\ref{newst}, right panel), and $\Mbh$ can increase up to a
factor of 4 (Fig.~\ref{mbh}, right panel), this implies that after
$z\sim 2$ the Magorrian relation is expected to shift upwards in
massive ETGs, just as a consequence of the passive evolution of the
stellar population. The extent of the shift is small, though, within
the scatter and uncertainties of the local and low redshift observed
relations (e.g. McConnell \& Ma 2013, Schulze \& Wisotzki 2014).

It is also interesting to compare $\DMstar$ with the total stellar
mass losses from the beginning to the end of the simulation, i.e. from
an age of 2 Gyr for the stellar population to that of 13 Gyr. From an
integration of the rate in eq. (1), these losses amount to $\sim 10$\%
of $\Mstar$. From Fig.~\ref{newst} we see that $\DMstar$ is $\simeq
5$\% of $\Mstar$ (or lower for the least massive models), thus,
roughly half of the integrated stellar mass losses within a galaxy
since $z\sim 2$ goes into new stars. The remaining part of the mass
losses goes for a minor fraction into $\DMbh$ (that reaches at most
$3\times 10^{-3}\Mstar$, and then at most 3\% of the integrated
stellar mass losses, see Fig.~\ref{mbh}), and for a major fraction is
ejected from the galaxy, due to SNe heating and the further help of
the AGN feedback action.  We recall that the AGN feedback produces
just an increase in the ejected mass from the galaxy, but has never
been found capable of producing a global/major outflow by itself
during the passive galaxy evolution after $z\sim 2$.  To better
quantify this increase, we can evaluate the amount of gas residing
outside a reference radius of 5$\re$ (that we consider lost by the
galaxy), at the end of the simulations, for models with and without
feedback. This amount is larger for feedback models than for NOF
models by a percentage ranging from $\sim 20$\% to $\sim 40$\%, from
the largest to the smallest galaxy masses.  We can also compute the
``load factor'', defined as the ratio between the gas ejected from the
galaxy during the evolution, and the mass in new stars ${\rm
  \DMstar}$; such a factor is useful to establish the possible role of
AGN feedback in adding material to the circumgalactic medium, after
$z\sim 2$.  For feedback models, the load factor is $\simeq 0.6$,
except for the lowest mass (E$^{180}$) galaxies, that eject a larger
amount of gas, and for which then the load factor is $\simeq 3$. For
NOF models, the factor is slightly larger ($\simeq 0.7$), but not so
different, because they eject less mass, but also form less stars; a
markedly lower factor ($\simeq 1$) than in feedback models is instead
shown by the E$^{180}$ NOF models, where the energy input by the
feedback can be significant in clearing the gas from the galaxy.  Note
that these load factors are similar to those ($\sim 0.7$) recently
determined thanks to background quasar lines of sight passing near
star-forming galaxies (e.g. Schroetter et al. 2016)

Figure~\ref{tm} shows the epoch at which half of $\DMstar$ is
formed ($t_M$ from Tab. 2). This epoch (measured since the birth of
the original stellar population, i.e., 2 Gyr before the start of the
simulation) keeps between 5 and 6 Gyr for most models, and tends to be
larger for MF than for FF models. Thus, half of the new stars created
during the life period of $2-13$ Gyr form quite early, during the
first $3-4$ Gyr.  Figure~\ref{f1} shows the time evolution of the SFR
for the E4$^{210}$ model, and Tab. 2 gives the instantaneous value of
the SFR at the end of the simulations. These present-epoch rates
typically keep below $\simeq 0.5\Msun$ yr$^{-1}$, and compare
well with the current rates observed for molecular gas-rich ETGs of
the local ATLAS$^{\rm 3D}$ sample, which range from $0.01$ to $3\Msun$ yr$^{-1}$, with a median value of $0.15\Msun$
yr$^{-1}$ (Davis et al. 2014).  Larger SFR than $\simeq 0.5\Msun$
yr$^{-1}$ are shown by models caught in an outburst (Fig.~\ref{f1}),
or in the most massive models, that host more gas, and have a larger
AGN activity; note that the ATLAS$^{\rm 3D}$ survey is not
representative of the most massive models in Tab. 2.

Another interesting aspect for a comparison with observations is {\it
  where} SF mainly takes place.  Figure~\ref{newsect} shows meridional
sections of the ratio between the density in newly formed stars, at
the end of the simulations, and that in the original stellar
population, for the E4 shape (results are similar for the E7 case).
SF is very low in the lower mass model (E4$^{180}$), that experiences
an almost global outflow over its whole lifetime.  SF is instead
evident in the larger mass E4$^{250}$ model: it forms a nuclear
stellar disk in the NOF case\footnote{This disk is the result of the
  stellar streaming imposed in the central regions for numerical
  reasons, and discussed in Sect. 4, see eq. (38).}, while it has a
roughly spherical distribution, peaking within $\simeq 1$ kpc radius,
in the cases with feedback. The extent of the SF is measured more
quantitatively by $r_M$, the radius including half of the new stars at
the end of the simulation (in Tab. 2). Typically $r_M\simeq \re /2$,
for feedback models, except for the least massive ones, where SF is
very low but extended. In contrast, $r_M$ is much smaller for NOF
models, indicating how most of SF takes place at the center, i.e., the
gas flows to the central regions before having time to start SF. In
feedback models, instead, as already noted, the gas is compressed from
time to time by AGN activity, and kept at larger distances by the
feedback action, which produces longer times available for SF, that is
then favored.  The feedback models also show an X-shaped structure in
the SF, inclined by $\simeq 45^{\circ}$, more evident in the MF
case. This is related with the aperture of the conical wind: at the
contact surface between the wind region and the external gas, the
density is increased, with a consequent increase in SF. Note that in
the simulations the newly formed stars remain in the position where
they are born, while in reality they move away from that place. From
this point of view, it is not clear whether these features could be
evident in real ETGs; however, curiously, X-shaped features have been
observed in the morphology of bulges (although on a much larger scale;
e.g. Ness \& Lang 2016), and in scattered light produced by
illuminated dusty cones in quasars hosting winds (e.g. Obied et
al. 2015).

When seen in the surface brightness profile, these new stars produce a
central cusp, as observed in ETGs at high angular resolution (see
Kormendy et al. 2009 for a review on these nuclear cusps or cuspy
cores). Even coreless galaxies do not have featureless power-law
profiles, but, rather, they show central extra light above the inward
extrapolation of the outer Sersic profile. Kormendy et al. (2009), and
Hopkins et al. (2009), suggested that the extra light is produced by
starbursts fed by gas dumped inward during dissipative mergers.  The
origin of the observed cusps could also reside in the evolutionary
phenomena investigated here (see also Ciotti \& Ostriker 2007, Ciotti
2009).

\subsection{The evolution of $\Mdotbh$, $\Lbh$ and the duty cycle}

Figure~\ref{lbh} shows the time evolution of $\Mdotbh$ (eq. 18), and
$\Lbh$ (eq. 23), that are relevant to quantify the accretion history
of the MBH, for the representative model E4$^{210}$ in
Fig.~\ref{f1}. In feedback models, the average accretion rate slowly
decreases with time, in pace with the declining rate of mass input
from stars.  The accretion episodes may extend throughout the galaxy
lifetime, with $\Mdotbh \ga 10^{-4}$M$_{\odot}$yr$^{-1}$, as for more
massive models, or for a shorter time, and with smaller $\Mdotbh$, as
for lower mass models (E$^{180}$).

A major difference with respect to 1D hydrodynamical simulations,
where major bursts are well separated in time (Ciotti \& Ostriker
2012), is that the bursting activity is here almost continuous (as
described in Section 5.1).  In spite of this difference between the
bursting activity of 1D and 2D hydrodynamics, the resulting duty-cycle
is not so different (both in spherical and flat galaxies). In order to
quantify this important aspect, we considered the time spent by each
model at any chosen Eddington ratio $l$, and with $l$ above/below a
chosen threshold.  In particular, for the E4 and E7 FF models,
Fig.~\ref{diff4} shows the fraction of total simulation time, and the
fraction of total radiative energy, respectively spent and emitted at
different values of $l$.  For all models the distribution of times has
a long, almost flat tail extending to the lowest $l<10^{-10}$, and
peaks at an $l$ that increases with the galaxy mass: the peak is
located at $l\simeq 10^{-3}$, and reaches $l\simeq 10^{-2}$ for the
most massive models (while becoming less and less pronounced).  All
distributions drop sharply above $l\simeq 10^{-2}$, so that all models
spend very little time above $l\simeq 10^{-1}$.  An average, realistic
galaxy, as could be described by the E4$^{210}$ FF model, spends 75\%
of its time since $z\sim 2.2$ below $l\approx 10^{-3}$.  Note however
how the fraction of the energy emitted at high Eddington ratios is
significant. Even though the nuclei are usually very faint, all models
typically emit 25\% of the energy above $l=0.1$, as shown by the right
column in Fig.~\ref{diff4}.

Figure~\ref{duty4} shows the percentage of the total simulation time
(11 Gyr), and of the total emitted energy, respectively spent and
emitted {\it above and below } each value of $l$, for the E4 FF models
(the same figure for the E7 FF models, not shown, presents the same
trends, both in shapes and in normalizations).  Estimates based on 1D
models, and with a luminosity threshold of $\Lbh=\Ledd/30$, gave a
figure of a few percent for the fraction of time spent above this
threshold (Ciotti et al. 2010). Remarkably, we can confirm this
estimate, with a percentage of time spent above $l=1/30$ ranging from
3 to 5\%, going from the E4$^{210}$ to the E4$^{300}$ models
(Fig.~\ref{duty4}, left panels, solid lines).

For reference, Table 2 gives the fraction of time spent above $l=0.05$
($\mathcal{D}$); the $l$-value with respect to which the MBH energy is
emitted equally above and below $l$ ($l_{0.5}$); and the time $t_L$ at
which half of the total MBH radiation energy, emitted over 2--13 Gyr,
has been emitted, measured since the birth of the galaxy (i.e., 2 Gyr
before the start of the simulation).

The duty-cycles resulting from the present set of simulations, in the
form shown in Figs.~\ref{diff4} and~\ref{duty4}, represent a useful
diagnostic that could be used in observational works, and
phenomenological models, that try to derive the growth rate of each
MBH mass, the evolution of their Eddington ratio, the total luminosity
emitted as a function of $l$, and the duty cycle of AGN activity
(e.g., Marconi et al. 2004, Trakhtenbrot \& Netzer 2012, Schulze et
a. 2015, Caplar et al. 2015). We stress that our predictions concern
massive non-rotating ETGs, with $\Mbh\ga 10^8$M$_{\odot}$, since
$z\sim 2.2$, i.e. after the galaxy formation has ended.  How much the
rotation of a galaxy could eventually affect these findings will be
addressed in a forthcoming paper.

\section{Summary and conclusions}\label{disc}

In this work we have followed the evolution of hot gas flows in ETGs
with central MBHs, using 2D hydrodynamical simulations where the most
accurate and physically consistent description of AGN feedback (both
radiative and mechanical, due to AGN winds) is implemented, the
spatial resolution at the center is parsec-scale, and the underlying
galaxy models are the most realistic dynamical models adopted so far
in this kind of studies. The mass and energy input from the stellar
population are secularly evolving, according to the prescriptions of
stellar evolution theory for the stellar mass losses, and to the
predictions of progenitors evolution and to observations, for the
declining SNIa's rate. Star formation is implemented via a simple
scheme based on physical arguments shown to reproduce well the
Kennicutt-Schmidt relation (Negri et al. 2015).  The galaxy models are
axisymmetric, and the Jeans equations provide the detailed internal
stellar kinematics on which the stellar kinematical heating is based.
The stellar density profile follows a deprojected (ellipsoidal) Sersic
law, and the main observables ($L,\re,\,\se$) are chosen in order for
the galaxy models to lie on the main scaling laws. The spherical dark
matter halo has a radial profile predicted by cosmological simulations
(NFW), and is normalized to account for a dark mass within $\re$ lower
than the stellar mass, as observed.  A few limitations remain in the
present work, though.  First of all, significant galactic rotation is
not included; rotation is known to lead to an almost corotating
gaseous halo (Negri et al. 2014b), which in turn produces a massive
rotating equatorial cold disk, of $\sim$kpc scale, that is expected to
be unstable, with the consequent disposal of fresh gas on the central
MBH. A new time-scale is then introduced, determined by the global
stability of the disk, and the amount of gas accreted on the MBH
depends also on the SF taking place in the disk.  Secondly, the
radiative transport is here calculated by using the simplified (and
computationally much faster) scheme described in Novak et al. (2012),
that has been shown to produce very accurate results when compared to
the exact numerical solution.  Also not included are all effects
related with the presence of dust (radiative transport, reprocessing
of the radiation from the new stellar population, sputtering, as
described in Hensley et al. 2014, Novak et al. 2012).  Third, the
circumnuclear sub-grid accretion disk, implemented in some of our
previous works (e.g. Ciotti \& Ostriker 2012) is not considered.
Finally, the presence of a jet, that seems relevant at recent epochs,
is still to be included.

The main results of this work are as follows.

1) $\Lx$ and $\Tx$ of the gas at the present epoch, for models with
feedback, reproduce well those observed for ETGs similar to our galaxy
models, at all galaxy masses; in particular, the observed large range
of $\Lx$ at the low $\Tx$ is accounted for, thanks to the prevalence
of outflows in lower mass galaxies. At fixed galaxy mass, a change in
the galaxy shape produces a lower $\Lx$ for flatter morphologies,
because the outflowing region becomes larger, as obtained for 2D
models without feedback (Negri et al. 2014b).  The evolution of the
gas when feedback is present is characterized by large and frequent
fluctuations in $\Lx$ and $\Tx$, except for the lowest mass models
explored here, where these fluctuations stop after the first few Gyr,
due to the onset of a global outflow sustained by SNIa's
heating\footnote{In these low mass ETGs the MBH accretes almost
  steadily, from a hot atmosphere, at a very sub-Eddington rate, as
  shown previously (e.g. Ciotti \& Ostriker 2012)}.  During the past
3--5 Gyr, the fraction of time spent above a value representative of
the largest gas emission observed for normal ETGs in the local
universe ($\Lx = 3\times 10^{41}$ erg s$^{-1}$), is $\la 1$\%.  After
each outburst, $\Lx$ within (1--2)$\re$ is lower by $\ga 1$ order of
magnitude than the same quantity for models without AGN feedback.

2) Each major burst triggered by accretion on the MBH is made by
several smaller bursts, and lasts $\sim$few$\times 10^7$ yr; we stress
that this timescale is not imposed a priori, but it results from the
simulations. The 2D hydrodynamics and the ellipsoidal shape of the
galaxy models produce a complicated gas evolution in the inner $\sim
2-3$ kpc region, where a cold and inflowing gas phase coexists with a
hot and outflowing one. Bursts from mechanical or full feedback are
qualitatively similar, but temporally more isolated in the purely
mechanical case. The X-ray surface brightness map shows a very bright
central region (on a scale of the order of $\sim$one hundred pc),
including sharp and very hot arcs, as a very transient feature. The
development and fading out of an outburst are also similar for
different galaxy shapes. We expect though that the addition of
galactic rotation will change this.

3) The mass in newly formed stars $\DMstar$, for feedback models,
ranges from $6\times 10^8$ to $3\times 10^{10}\Msun$, from the
low to the high galaxy masses. Since the models are basically
non-rotating, SF takes place in an almost spherical region; the radius
including half of the new stars is $r_M\simeq \re /2$, except for the
least massive models, where SF is very low but extended.  Half of the
mass in the new stars forms within 5--6 Gyr from the birth of the
original stellar population.

4) At fixed $\Mstar$, $\DMstar$ is larger for the full feedback case
than the mechanical one, that in turn produces a larger $\DMstar$ than
without feedback. This proves that feedback has the overall effect of
making SF more efficient, at least in the phase following the major
galaxy formation period studied here, when SF and MBH accretion are
fueled by passive stellar evolution. These results confirm previous
findings of spherical models about {\it positive feedback} (Ciotti \&
Ostriker 2007; e.g. see also Nayakshin \& Zubovas 2012, Ishibashi \&
Fabian 2012); thus geometrical effects allowed for by 2D simulations
and a flat galaxy shape do not invalidate the global picture.

5) $\DMstar/\Mstar$ is of the order of 0.04$-$0.05, excluding the
lowest mass models that show lower values ($\simeq 0.01$), since most
of their injected stellar mass loss is expelled in an outflow. On
average, excluding the lowest mass models, roughly half of the total
mass losses since an age of $\simeq 2$ Gyr ends recycled into new
stars.  The other half goes for a minor fraction into $\DMbh$ (for
feedback models), and for a major fraction is ejected from the galaxy,
mostly due to SNe heating.  AGN feedback produces just an increase (of
the order of $20-40$\%) in the ejected mass from the galaxy, during
the passive galaxy evolution after $z\sim 2$.  The ``load factor'',
defined as the ratio between the gas displaced outer of 5$\re$ during
the evolution, and the mass in new stars $\DMstar$, is $\simeq 0.6$
for feedback models, and slightly larger for NOF models (due to their
lower ability to form new stars). The lowest mass (E$^{180}$) galaxies
eject a larger amount of gas, and for them the load factor is larger
($\simeq 3$ for models with AGN feedback).

6) Final $\Mbh$ values of feedback models are consistent with those
measured in the local universe, at all $\Mstar$ considered here, while
they are overwhelmingly large without feedback; the latter is then
fundamental to keep the MBH masses at the values observed, {\it after}
$z\sim 2$.  In feedback models just $\simeq 3$\% of the total injected
mass from stars is accreted on the MBH; the average increase in $\Mbh$
since $z\sim 2$ is just of a factor of few ($2-4$).  The MBHs in more
massive galaxies grow more than in lower mass ones, during the later
phase of (small) MBH growth studied here. At fixed $\Mstar$, flatter
models show lower $\Mbh$, due to the lower gas binding energy, and FF
models end with lower $\Mbh$ than MF ones, thanks to their larger
capability of preventing gas accretion. $\DMbh/\DMstar$ ranges from
$8\times 10^{-3}$ to $7\times 10^{-2}$, from the lowest to the largest
galaxy masses. The implementation of rotation could change this trend,
by making the MBH growth less favored, and SF in the cold disk more
favored.

7) Half of the radiative output from the AGN is emitted within
$t_L\approx 5$ Gyr since the birth of the galaxy, roughly the same
value of the timescale $t_M$ within which half of the new stars form;
this is another indication that the two activities are linked. We
notice a slight but clear dependence on galaxy mass in the time-shift
between these two timescales: $t_L$ precedes $t_M$ in lower mass
galaxies, while the reverse is true in larger mass ones.

8) The outburst activity is almost continuous, and extends to the
present epoch, for more massive galaxies, or stops a few Gyr ago, for
the least massive ones considered here.  Almost all of the time is
spent at very low nuclear luminosities (75\% of the simulation time is
spent at Eddington ratios $l\la 10^{-3}$), but the energy is emitted
at high Eddington ratios, with one quarter of the total energy emitted
at $\Lbh/\Ledd>0.1$.  The resulting duty-cycles of AGN activity,
estimated as the fraction of time spent above $\Ledd/30$, ranges
from 3 to 5\%.

We note that recently Cheung et al. (2016) identified a population of
local ETGs (that they called “red geysers”) showing bisymmetric
outflow-like structures in their ionized gas emission line maps. The
authors argue that these are low-luminosity AGN-driven winds, and that
they are occurring in roughly 5-10\% of the red sequence at moderate
masses.  This finding would prove how the AGN wind mechanical feedback
continues down to late times, as predicted by our modeling.

\begin{acknowledgements}

  We thank G. Novak for providing an initial version of the code used
  here, and S. Posacki for providing the dynamical properties of the
  underlying galaxy models.  LC, SP and AN were supported by the MIUR
  grant PRIN 2010-2011, project ``The Chemical and Dynamical Evolution
  of the Milky Way and Local Group Galaxies'', prot. 2010LY5N2T.

\end{acknowledgements}

\newpage

\begin{figure}
\hskip 2truecm
\includegraphics[scale=0.65]{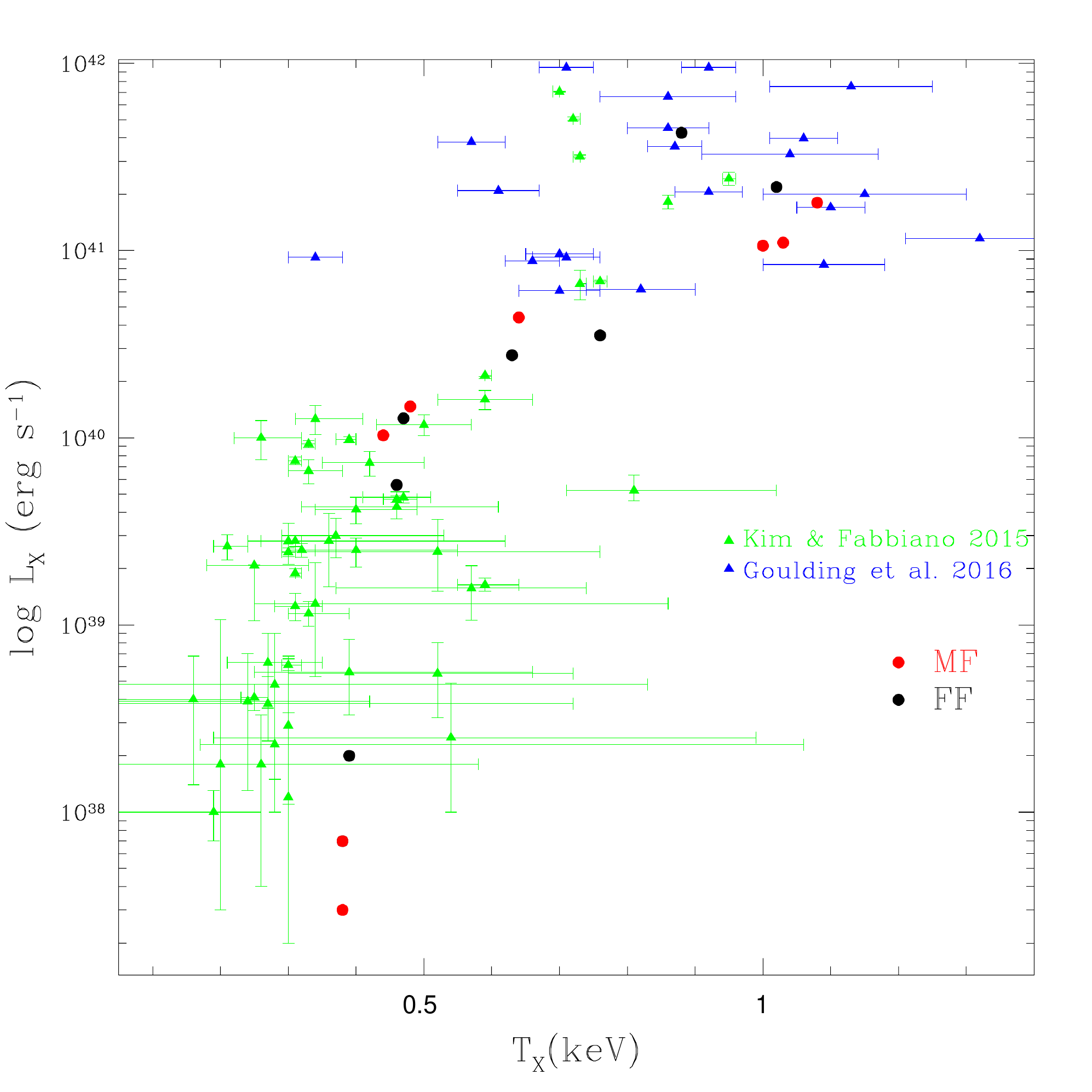}   
\caption{Relation between hot gas luminosity $\Lx$ and temperature $\Tx$, calculated at the present epoch and over the whole grid, 
for feedback models in Tab. 2;
$\Lx$ and $\Tx$ coincide with those in Tab. 2 (except for the lowest mass galaxies).
Red and black symbols are MF and FF models.
Also shown with errorbars are hot gas luminosities and temperatures for observed ETGs, recently calculated using $Chandra$ data
for the whole emission region, and after subtraction of resolved point-sources and
unresolved stellar emission (from Kim \& Fabbiano 2015, in green, and Goulding et al. 2016, in blue).
Most ETGs with $\Lx>10^{41}$ erg s$^{-1}$ are central galaxies in groups.}
\label{f2} 
\end{figure}

\newpage

\begin{figure*}
\vskip -6truecm
\hskip -4truecm
\includegraphics[scale=0.85,angle=-90]{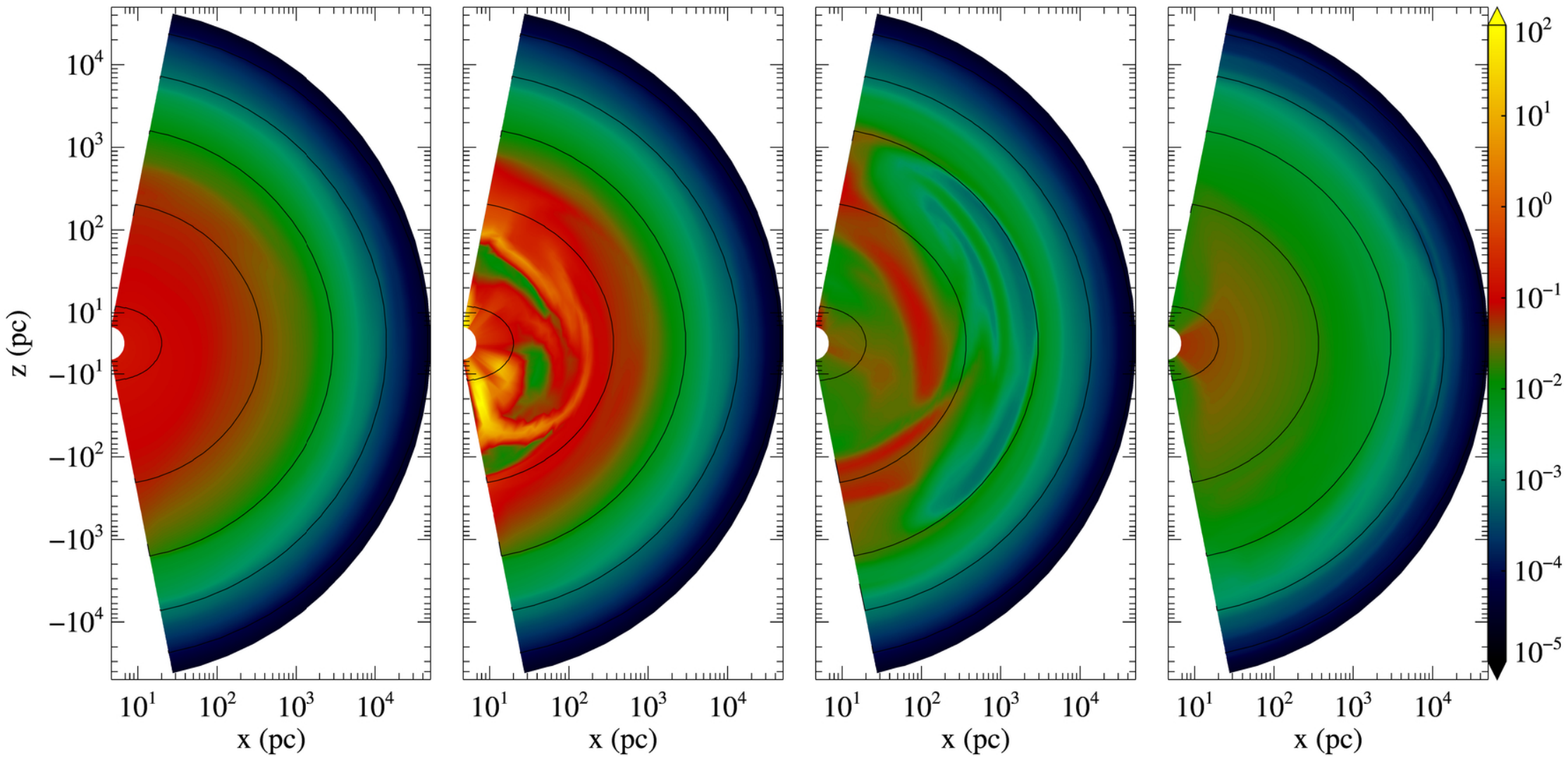} 
\vskip -2truecm
\caption{Map of the gas density for the E4$^{180}$ FF
 model, at four representative times in correspondance of an outburst at 6.85 Gyr: from
left to right t=6.84 Gyr (smooth appearance, immediately before the outburst), t=6.85 Gyr
(close to the peak in emission; cold and dense fingers are approaching the galactic center, where
outflow and inflow regions coexist; these are mixed with hot and low density regions already created 
by the outburst; see the next Fig. 4 for the mentioned features in the temperature, and Fig. 5 for those in the velocity), t=6.86 Gyr 
(the outburst is fading), and t=6.95 Gyr (the main outburst effects have vanished, the galaxy is left with less dense gas 
than in the leftmost panel, before the outburst; a less dense bi-conical region close to the nucleus is evident, produced by the AGN wind). 
The solid lines represent the optical isophotes.
}
\label{dens} 
\end{figure*}

\vskip -3truecm
\begin{figure*}
\vskip -2truecm
\hskip -1.5truecm
\includegraphics[scale=0.65,angle=-90]{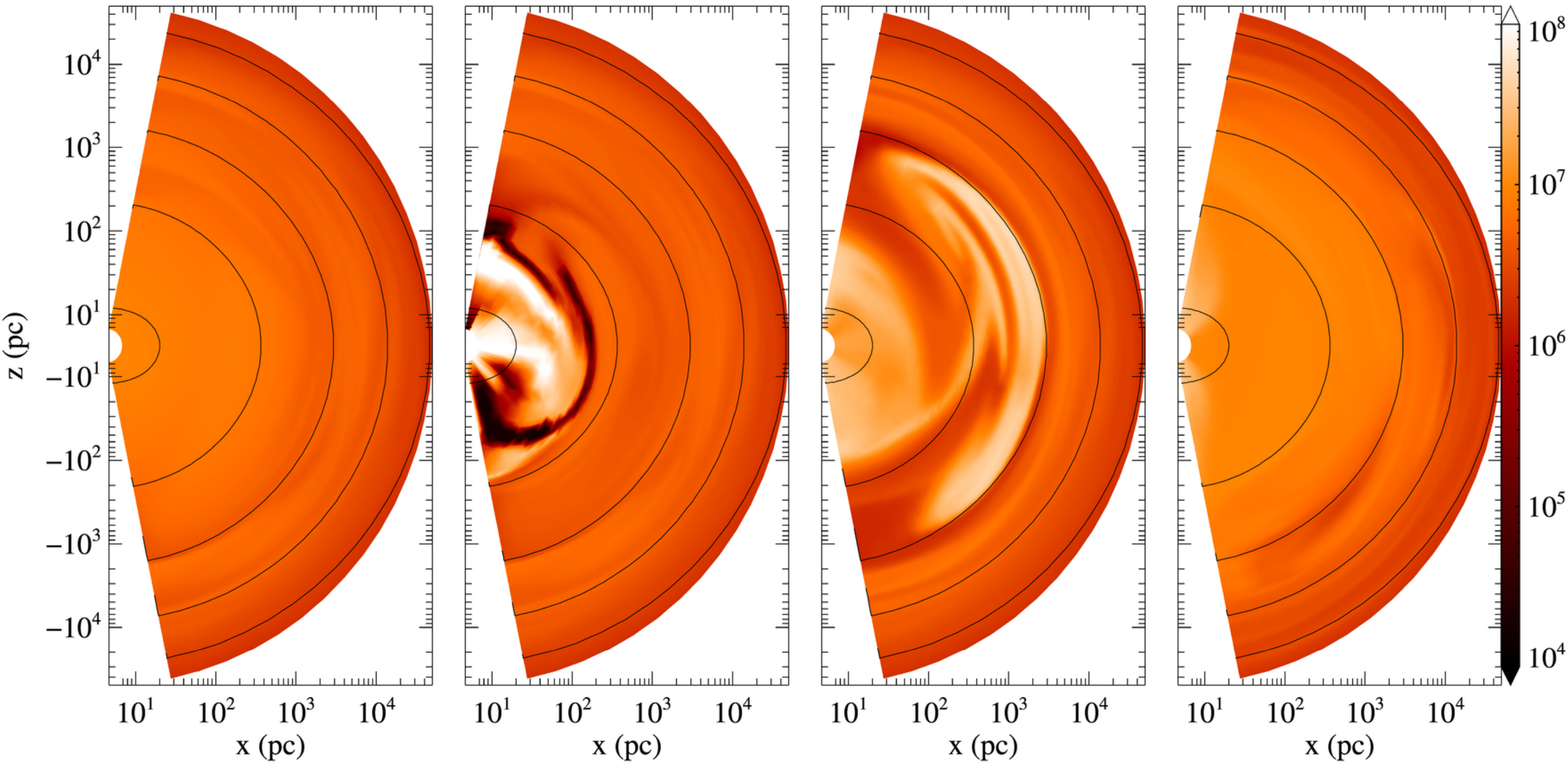} 
\vskip -2truecm
\caption{Map of the gas temperature for the same E4$^{180}$ FF
 model during the outburst at 6.85 Gyr of the previous Fig. 3, at the same representative times.
From left to right t=6.84 Gyr (smooth appearance, immediately before the outburst), t=6.85 Gyr
(close to the peak in emission; cold fingers are approaching the galactic center, while very hot regions 
have already been created by the outburst), t=6.86 Gyr (the outburst is fading; there is still some 
hot material outflowing at a radius of $\sim 1$ kpc), and t=6.95 Gyr (the main outburst effects have vanished;
the heating effect of the conical wind of the fading AGN is visible as an inner slightly hotter 
bi-conical region).
}
\label{temp} 
\end{figure*}

\newpage

\begin{figure*}
\vskip -2truecm
\hskip -1.5truecm
\includegraphics[scale=0.65,angle=-90]{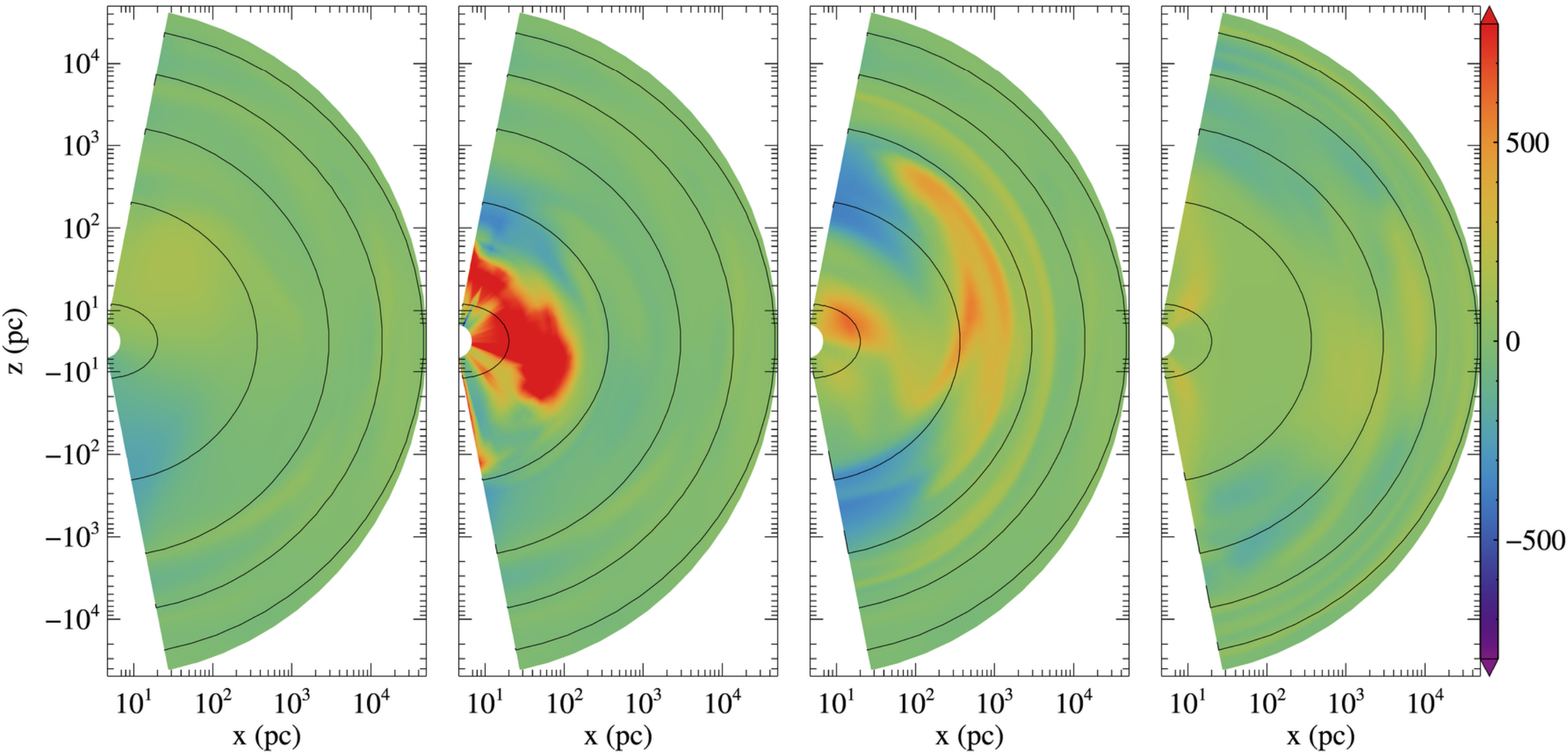} 
\vskip -2truecm
\caption{The same as in the previous Figs. 3 and 4, for the map of the radial component ($u_{\rm r}$) of the gas velocity.
Redder regions are outflows, bluer regions indicate inflows.
From left to right t=6.84 Gyr (immediately before the outburst), t=6.85 Gyr
[close to the peak in emission; outflow (hot) and inflow (cold) regions coexist around the galactic center], 
t=6.86 Gyr (the outburst is fading; there is still some outflowing gas from the center), and 
t=6.95 Gyr (the main outburst effects have vanished, the conical wind of the fading AGN becomes 
visible as an hourglass feature at the nucleus).}
\label{vr} 
\end{figure*}

\begin{figure*}
\vskip -2truecm
\hskip -1.5truecm
\includegraphics[scale=0.65,angle=-90]{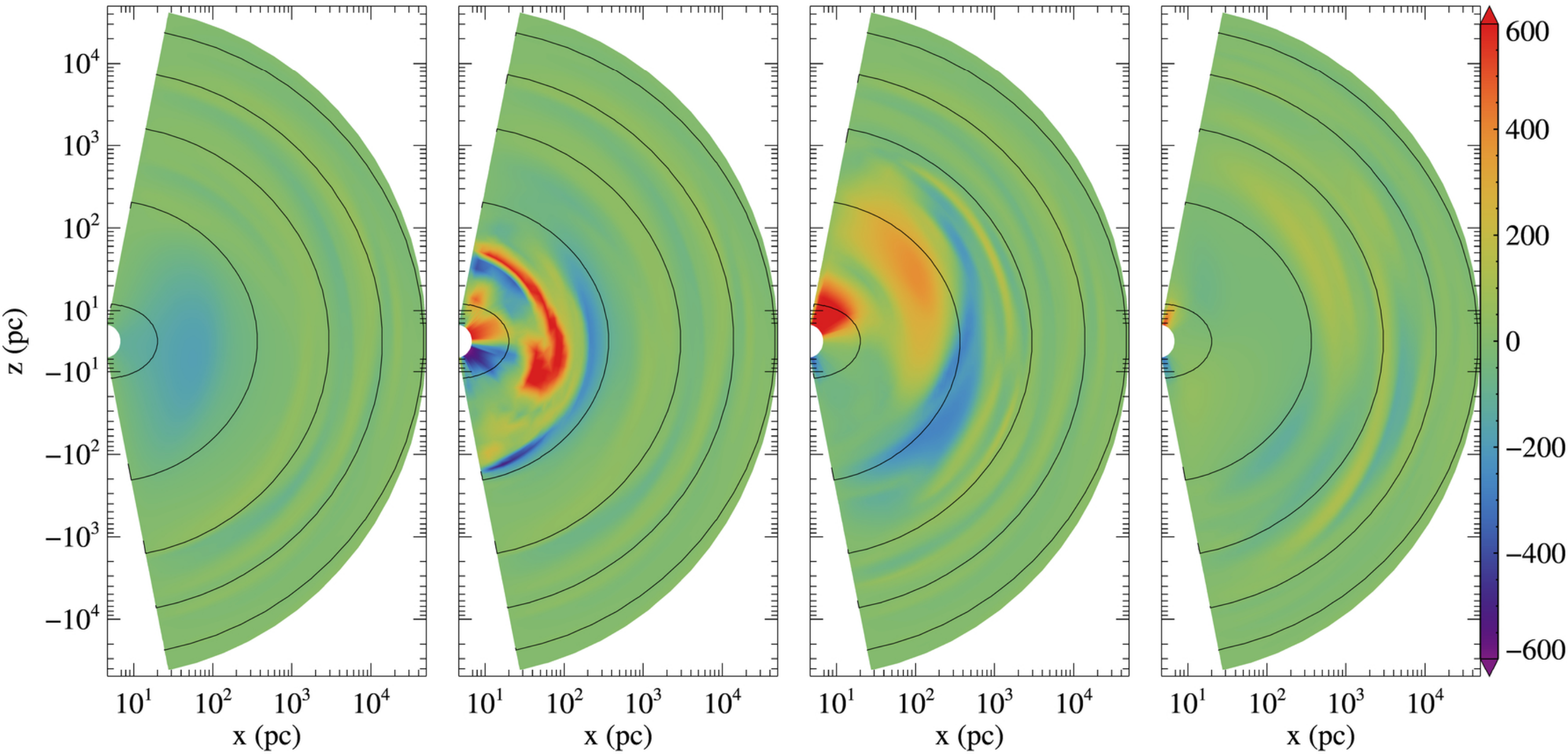} 
\vskip -2truecm
\caption{Map of the meridional tangential component ($u_{\rm {\theta}}$)  of the gas velocity for the outburst of the previous Figs. 3--5. 
Redder regions are moving clockwise, bluer ones are moving counter-clockwise.
From left to right t=6.84 Gyr (smooth appearance, immediately before the outburst), t=6.85 Gyr
(close to the peak in emission), t=6.86 Gyr (the outburst is fading), and 
t=6.95 Gyr (the main outburst effects have vanished; note close to the nucleus the tangential motions imparted to the flow by the AGN 
conical wind).}
\label{vt} 
\end{figure*}

\newpage

\begin{figure*}
\vskip -3.5truecm
\hskip -1.5truecm
\includegraphics[scale=0.65,angle=-90]{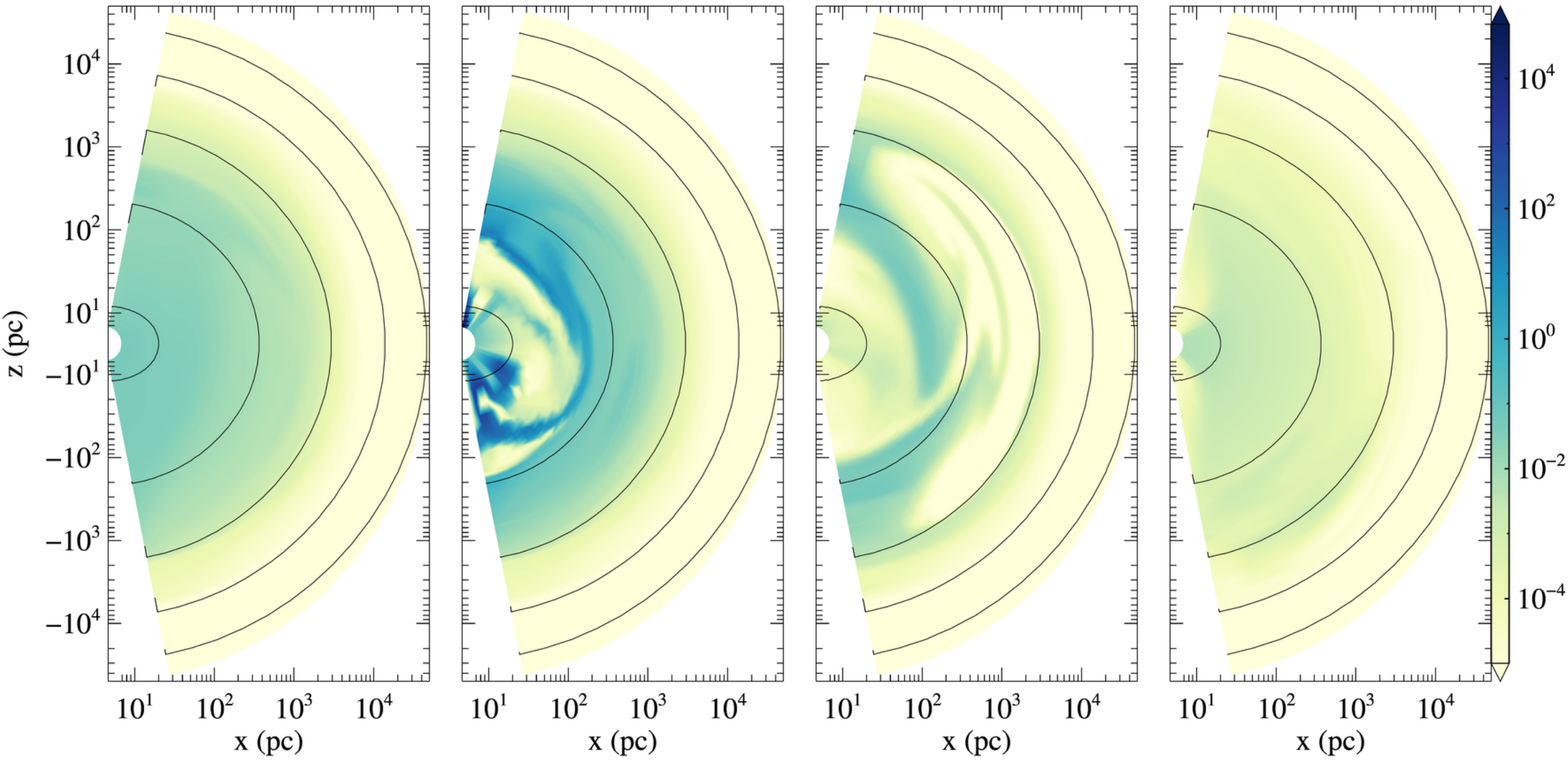} 
\vskip -2truecm
\caption{Map of the star formation rate density for the E4$^{180}$ FF
model during the outburst of the previous Figs. 3--6.
From left to right t=6.84 Gyr (immediately before the outburst), t=6.85 Gyr
(close to the peak in emission), t=6.86 Gyr (the outburst is fading), and 
t=6.95 Gyr (the main outburst effects have vanished).}
\label{sfrd} 
\end{figure*}

\begin{figure*}
\includegraphics[scale=0.4,angle=-90]{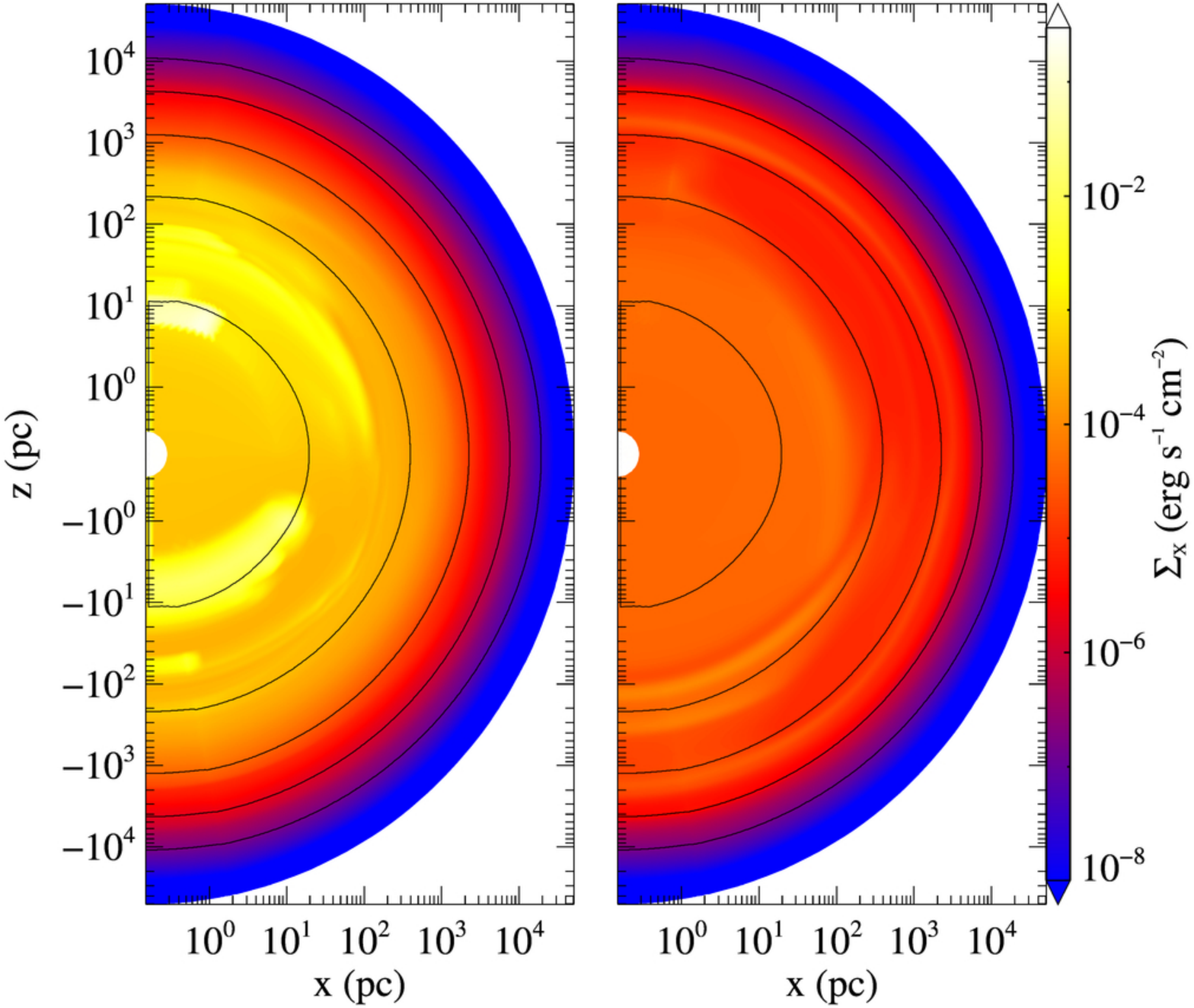}\\
\includegraphics[scale=0.4,angle=-90]{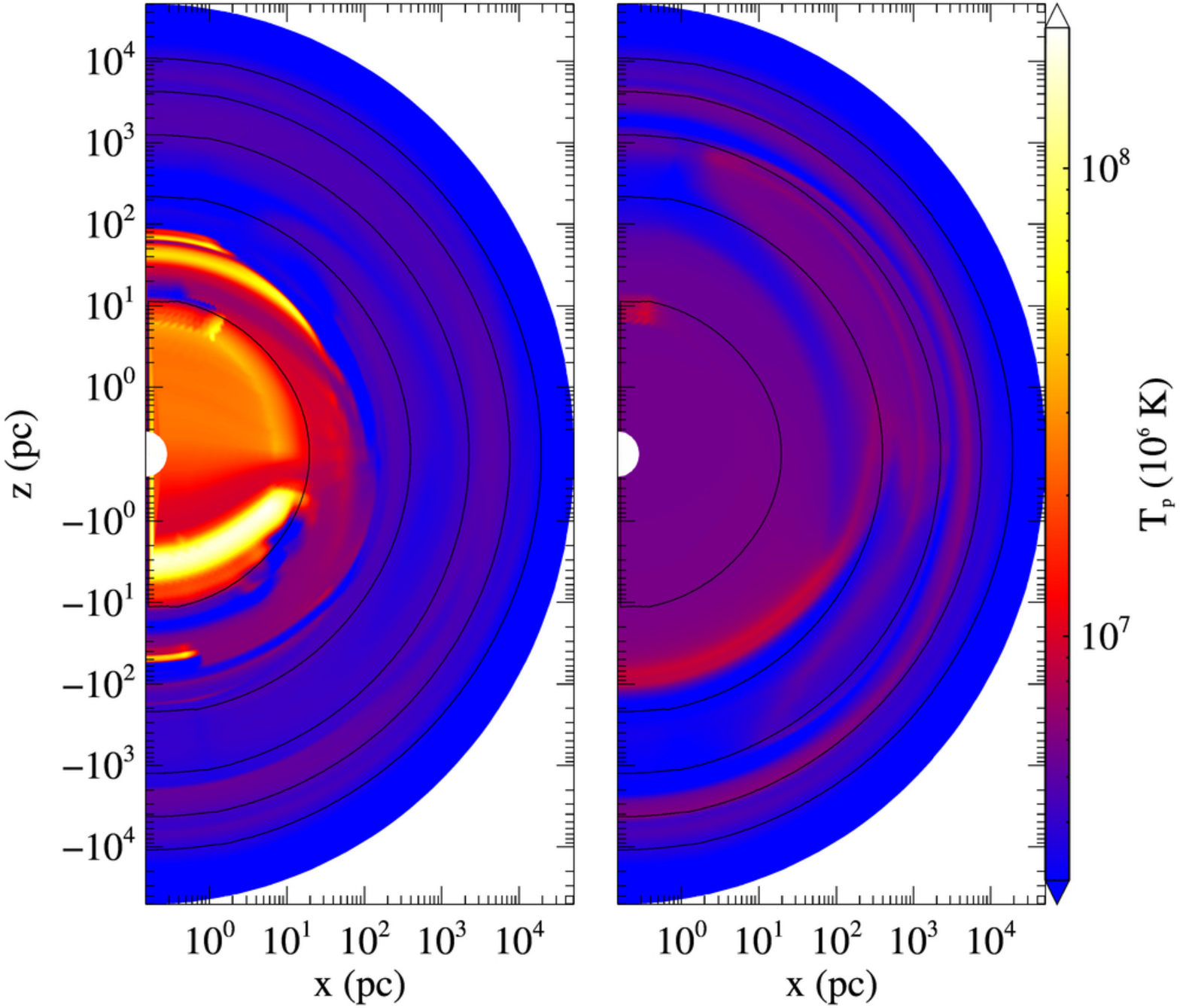}\\
\caption{Surface brightness maps in the 0.3--8 keV band (upper panels),
and projected temperature maps (lower panels), for the same model in outburst of the 
previous Figs. 3--7, at the two central times in those panels:  t=6.85 Gyr (the outburst has just 
started), and t=6.86 Gyr (the outburst is fading). Solid contours indicate the optical isophotes.
}
\label{brill}
\end{figure*}

\begin{figure*}
\includegraphics[scale=0.9]{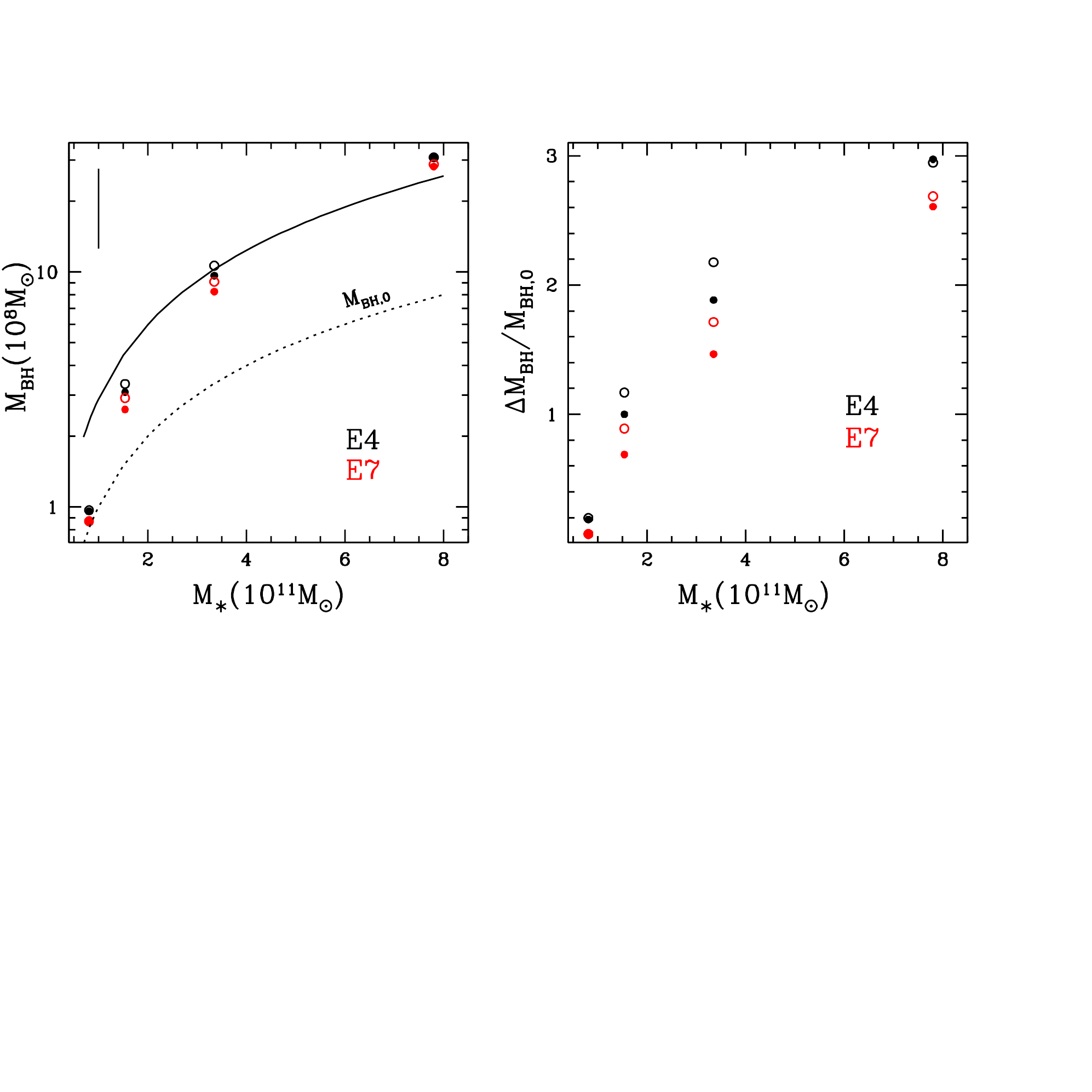}       
\vskip -8truecm
\caption{Left panel: final values of the MBH masses versus $\Mstar$ (given in Tab. 1). 
Plotted are the 16 models with feedback in Tab. 2: E4 in black, E7 in red;
open symbols are MF models, full symbols are FF ones. The positions of the models are compared with the
initial ${\rm M_{BH,0}-\Mstar}$ relation (dotted line), and the observed $\Mbh-\Mstar$ 
relation (solid line), derived for dynamically measured stellar masses of local ETGs (McConnell \& Ma 2013). 
The vertical bar shows the estimated intrinsic scatter in
log$\log\Mbh$ for the plotted scaling relation.
Right panel: the percental increase in the MBH mass, for the same
feedback models on the left, with the same meaning of symbols.}
\label{mbh} 
\end{figure*}

\begin{figure*}
\hskip 0.2truecm
\vskip -1truecm
\includegraphics[scale=0.9]{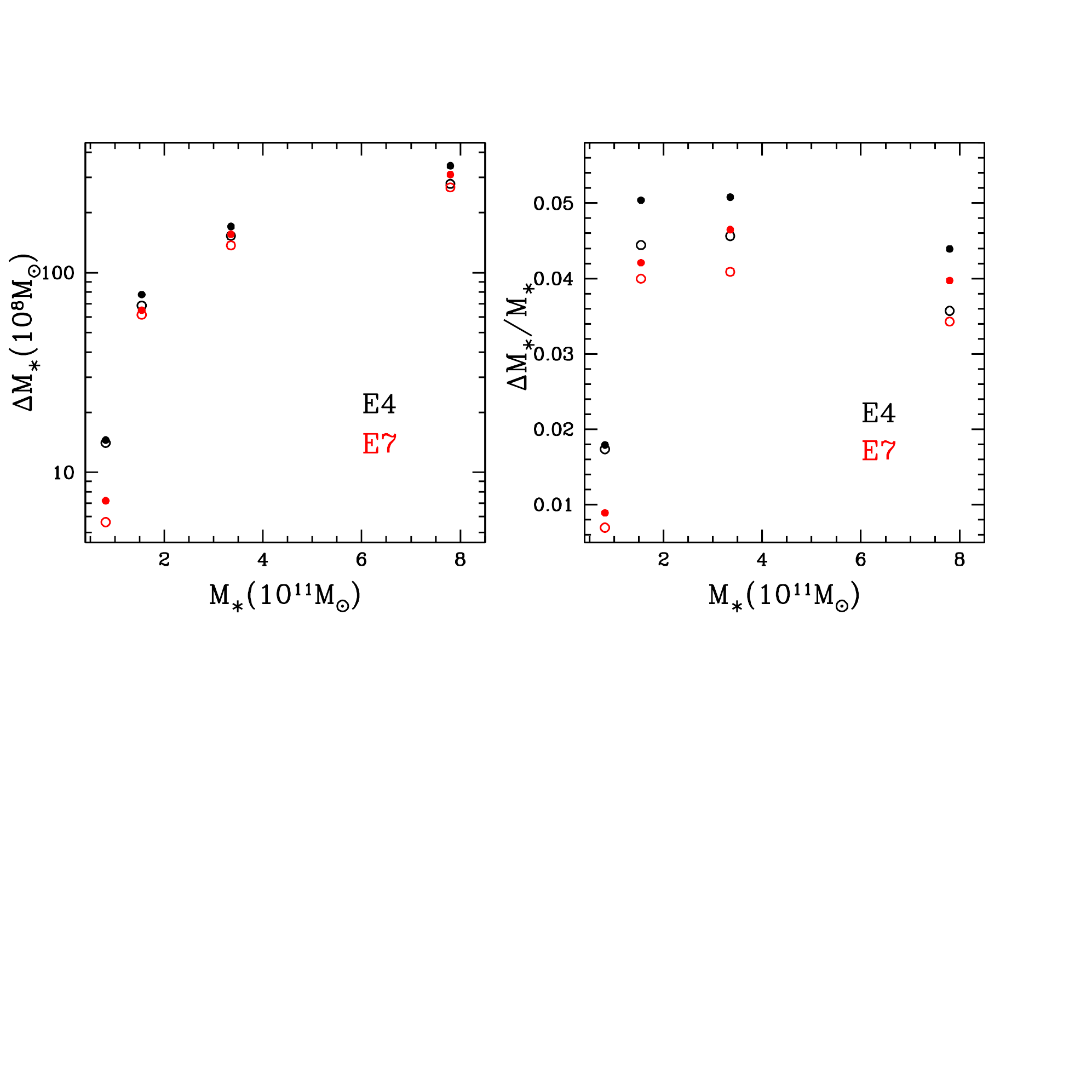}       
\vskip -8truecm
\caption{Left panel: final mass in the newly formed stars $\DMstar$ (from Tab. 2) versus
$\Mstar$ (from Tab. 1), for models with feedback (E4 in black, E7 in red).
Plotted are the 16 models with feedback in Tab. 2; MF and FF models are shown with open and full symbols respectively.
Right panel: the percental increase in the stellar mass, for the same models on the left, with the same meaning of
the symbols.}
\label{newst} 
\end{figure*}

\begin{figure}
\hskip 0.2truecm
\vskip -1truecm
\includegraphics[scale=0.9]{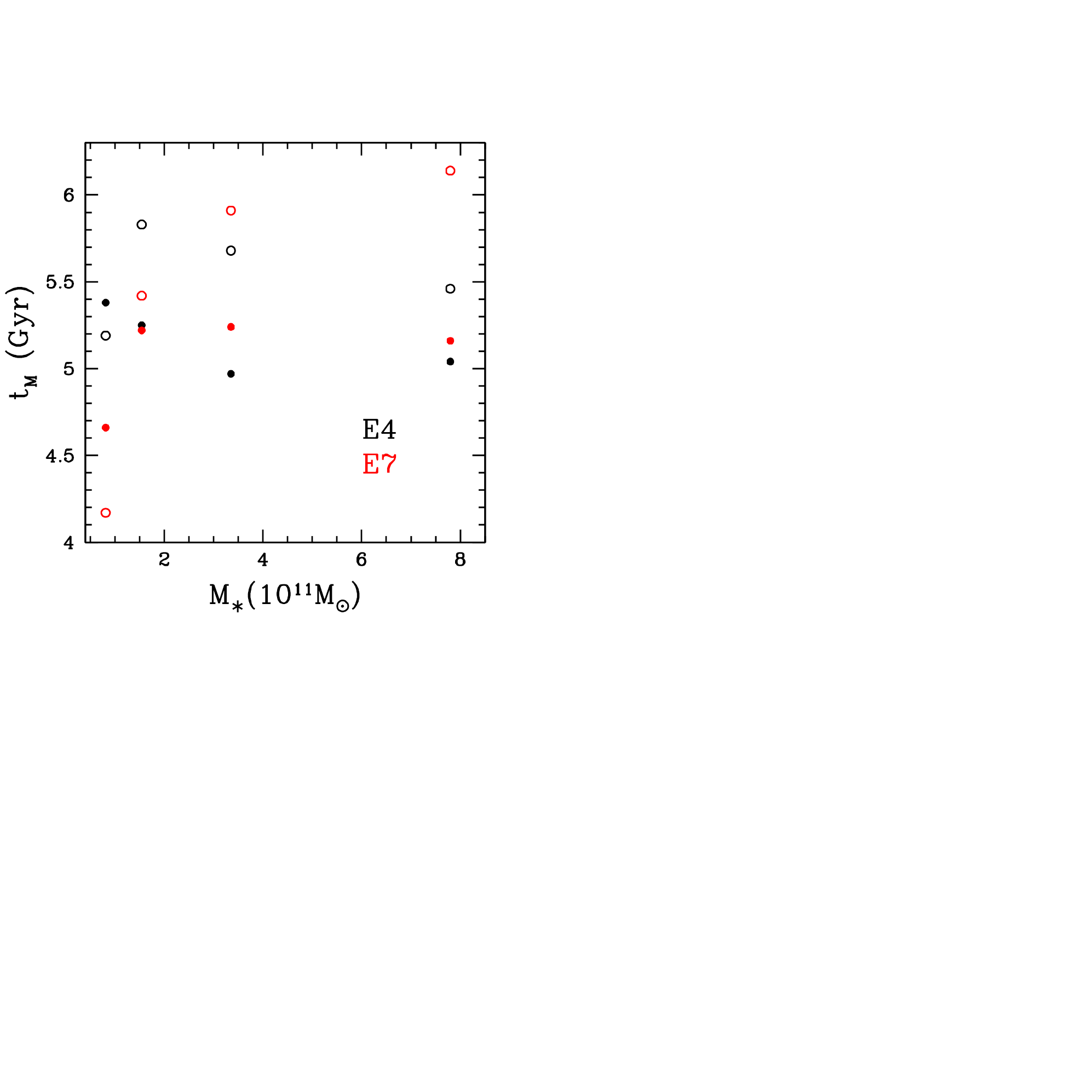}       
\vskip -8truecm
\caption{Epoch (measured since the birth of the original stellar
  population of the galaxy, i.e. 2 Gyr before the start of the
  simulation) at which half of the final mass in the newly formed
  stars $\DMstar$ is formed ($t_M$ in Tab. 2), versus $\Mstar$ (in
  Tab. 1); E4 models are plotted in black, E7 in red.  Plotted are the
  16 feedback models in Tab. 2; MF and FF models are shown with open
  and full symbols respectively.}
\label{tm} 
\end{figure}

\begin{figure*}
\vskip -1.truecm
\hskip -1truecm
\includegraphics[scale=0.65,angle=-90]{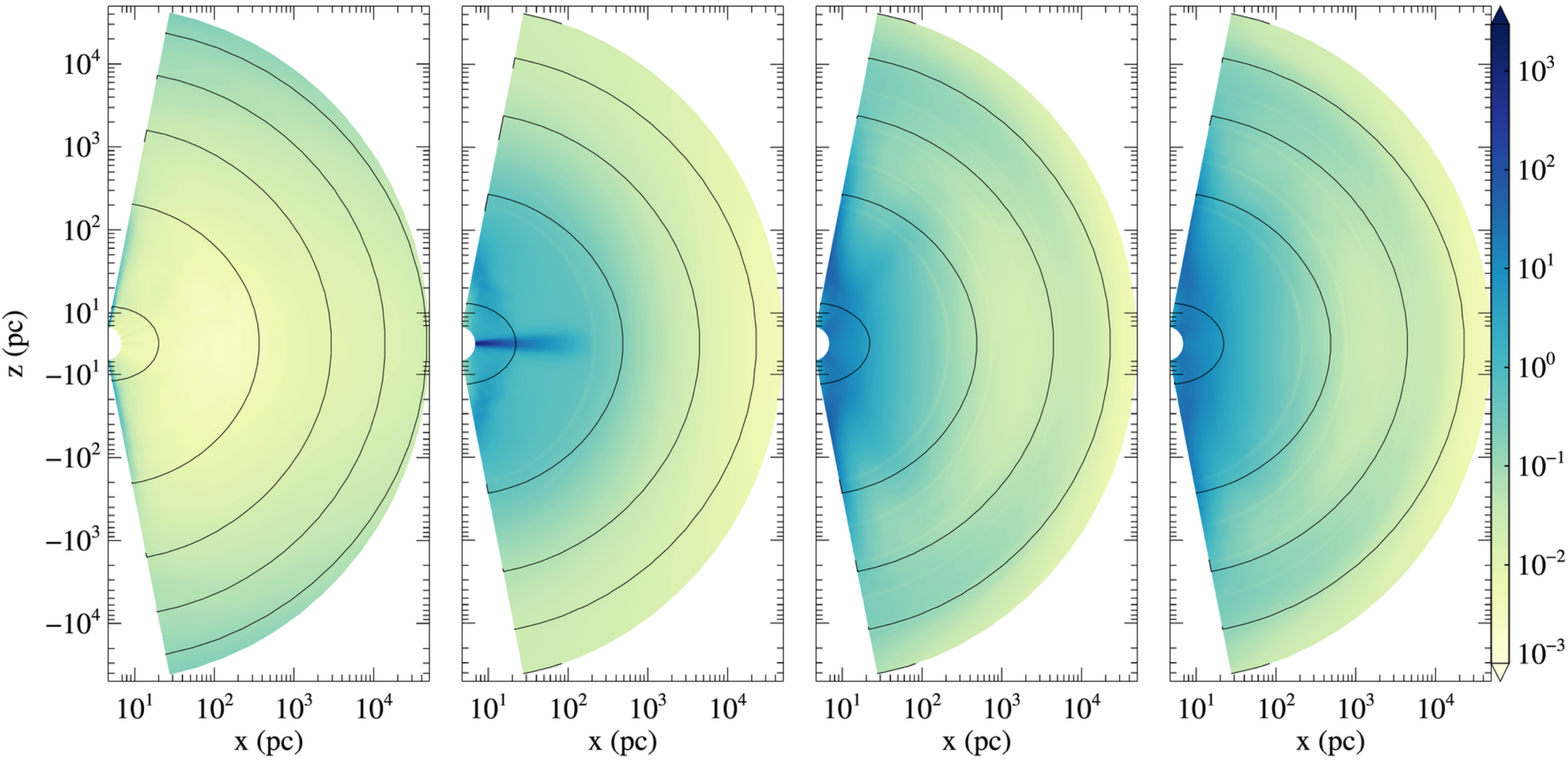} 
\vskip -1.5truecm
\caption{Meridional sections of the ratio between the density in newly
  formed stars at the end of the simulations and that in the original
  stellar population. Solid lines show constant densities for the
  original stellar population. From left to right the panels refer to
  the FF E4$^{180}$ model, and to the NOF, MF and FF E4$^{250}$
  models. SF is very low in the lower mass model, that experiences an
  almost global outflow over its whole lifetime; SF is instead
  significant in the larger mass galaxy, and it forms a nuclear
  stellar disk in the NOF case, while it has a roughly spherical
  distribution in the cases with feedback (see Sect. 5.3 for more
  details).}
\label{newsect} 
\end{figure*}

\begin{figure*}
\vskip -10truecm
\includegraphics[scale=0.9]{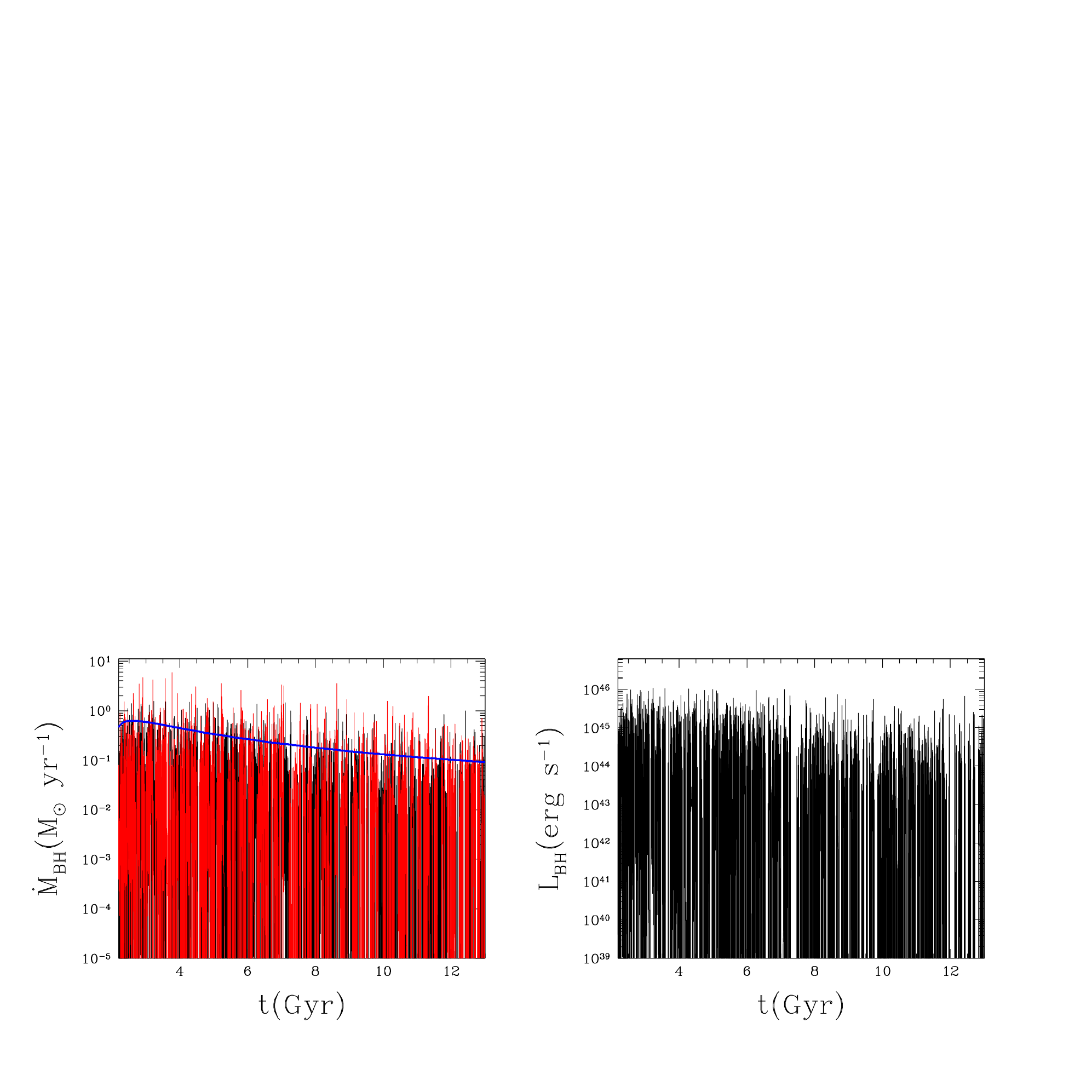}\\         
\vskip -12truecm
\includegraphics[scale=0.9]{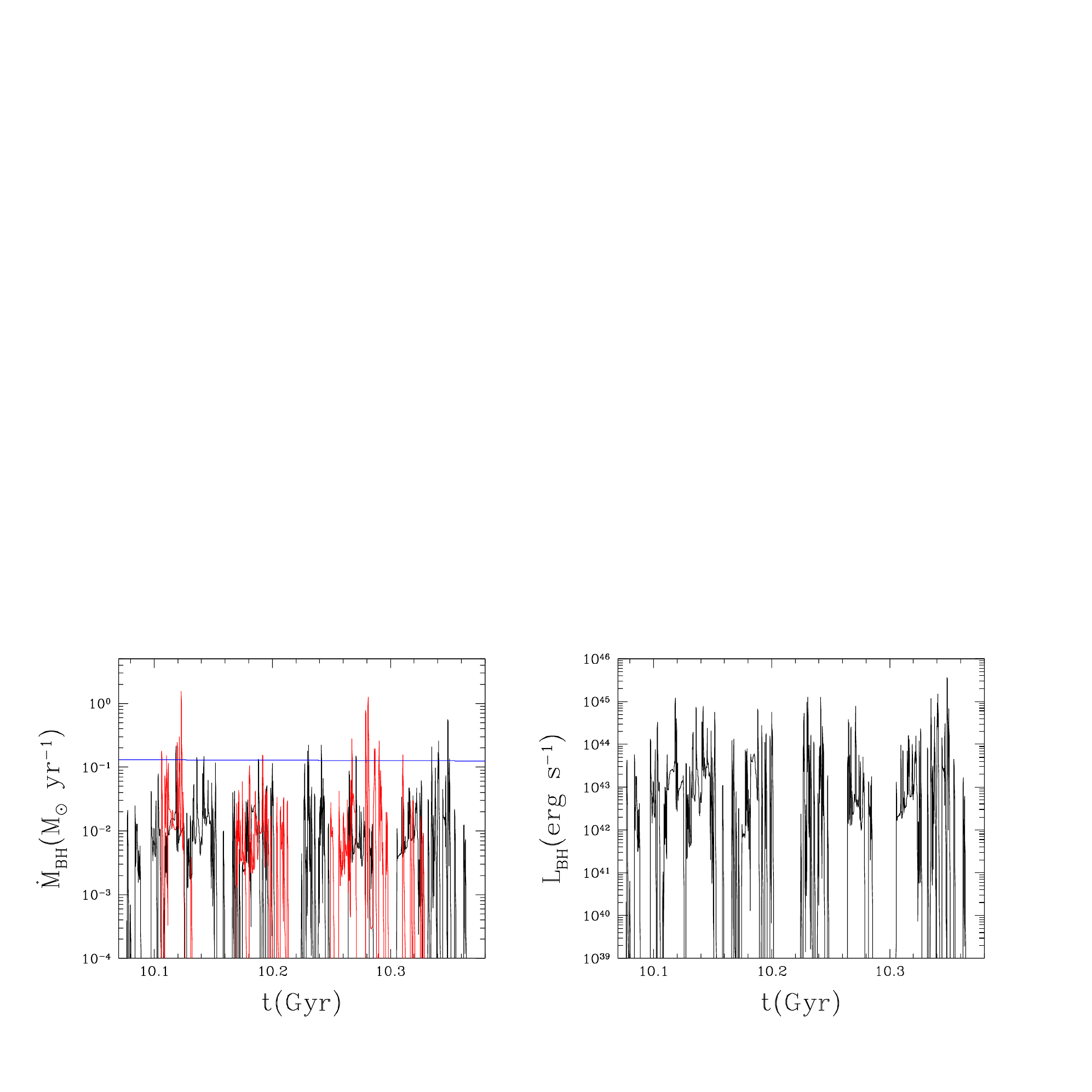}        
\vskip -1truecm
\caption{Time evolution of the mass accretion rate on the MBH ($\Mdotbh$, left), and the radiated accretion luminosity
($\Lbh$, right), for the E4$^{210}$ model (for which Fig. 1 shows the gas evolution).
Blue, red and black lines  refer to NOF, MF and FF models; in the right panel the nuclear luminosity is shown only in the FF case.
The lower panels show the same zoom in time as in Fig. 1. }
\label{lbh} 
\end{figure*}

\begin{figure*}
\hskip 2truecm
\includegraphics[scale=0.95]{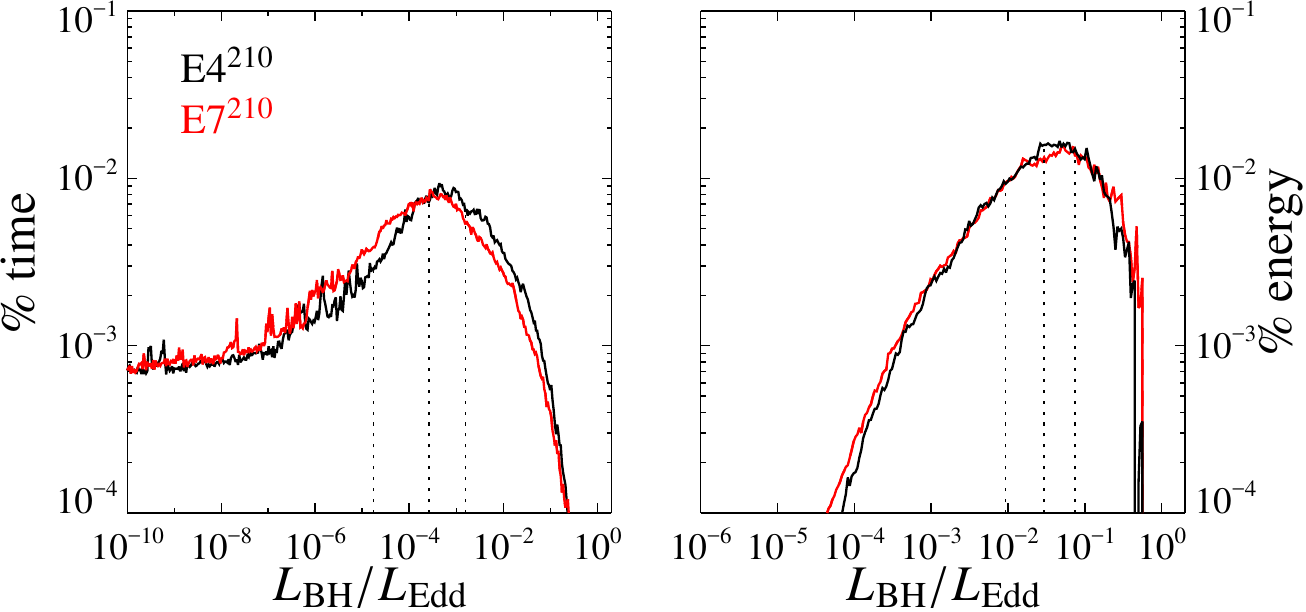}\\
\vskip 1truecm 
\hskip 2truecm
\includegraphics[scale=0.95]{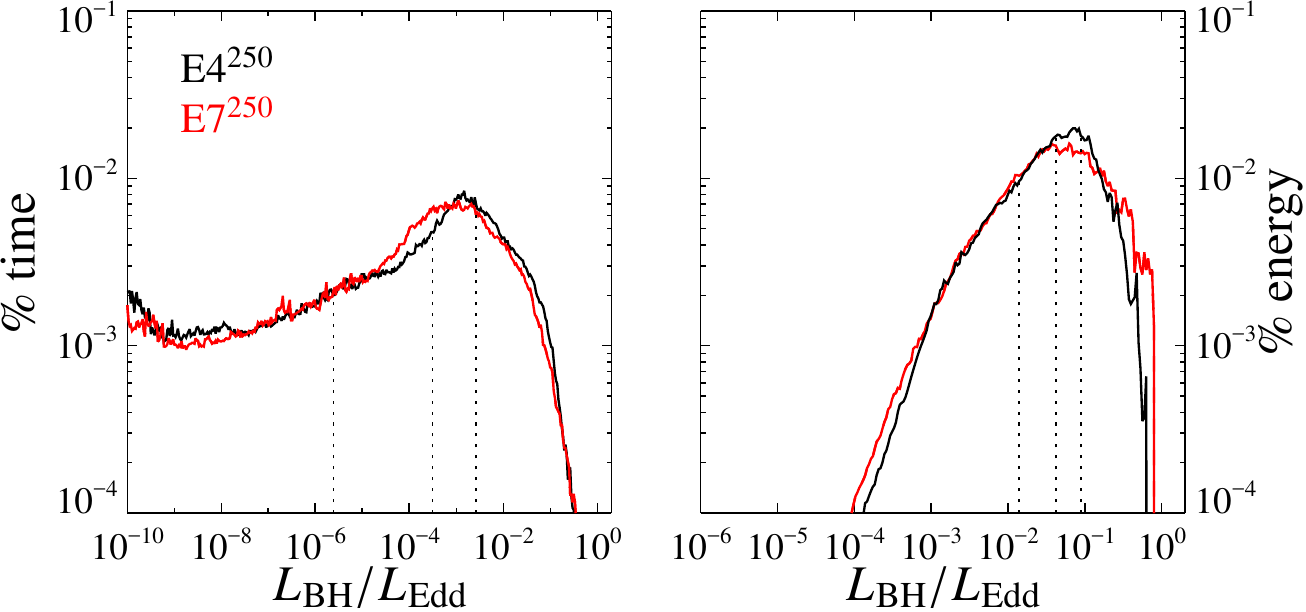}\\
\vskip 1truecm 
\hskip 2truecm
\includegraphics[scale=0.95]{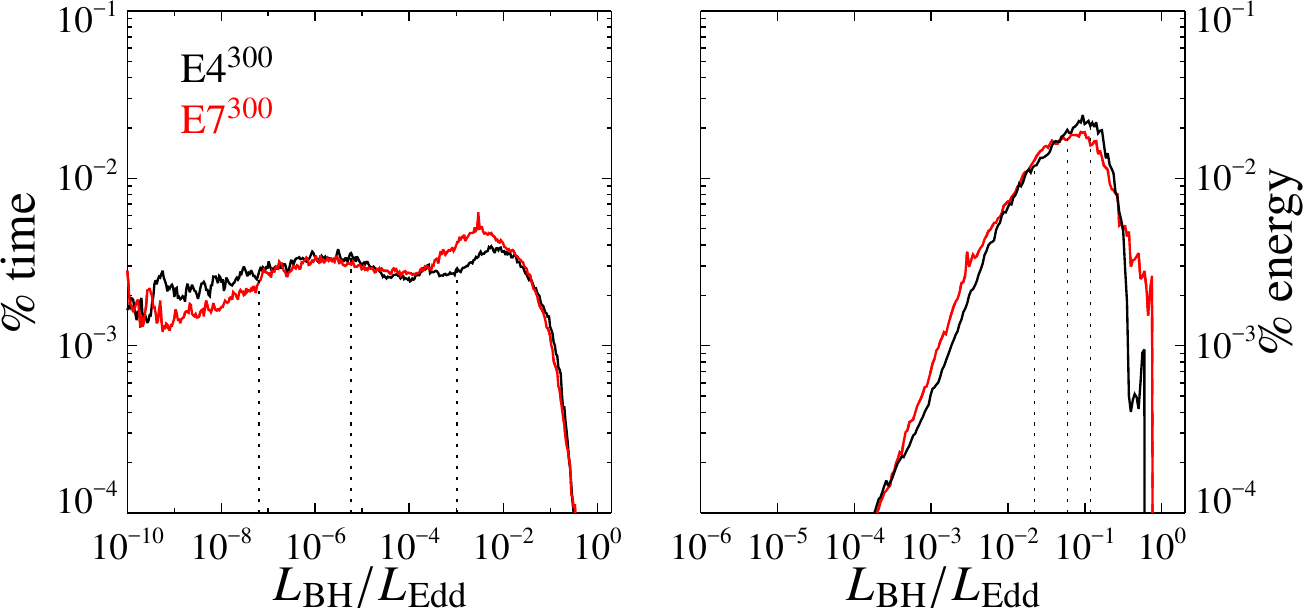}\\
\caption{Left panels: percentage of the total simulation time (11 Gyr)
  spent at the values of the Eddington ratio on the x-axis, for the E4
  FF models.  Right panels: for the same models on the left,
  percentage of the total energy emitted at the Eddington ratio on the
  x-axis.  In each panel, for the E4 models only, the vertical dashed
  lines mark the Eddington ratios below which the model spends 25\%,
  50\% and 75\% of the total time (left panels), or below which 25\%,
  50\% and 75\% of the total energy is emitted (right panels).}
\label{diff4} 
\end{figure*}

\begin{figure*}
\hskip 2truecm\includegraphics[scale=0.9]{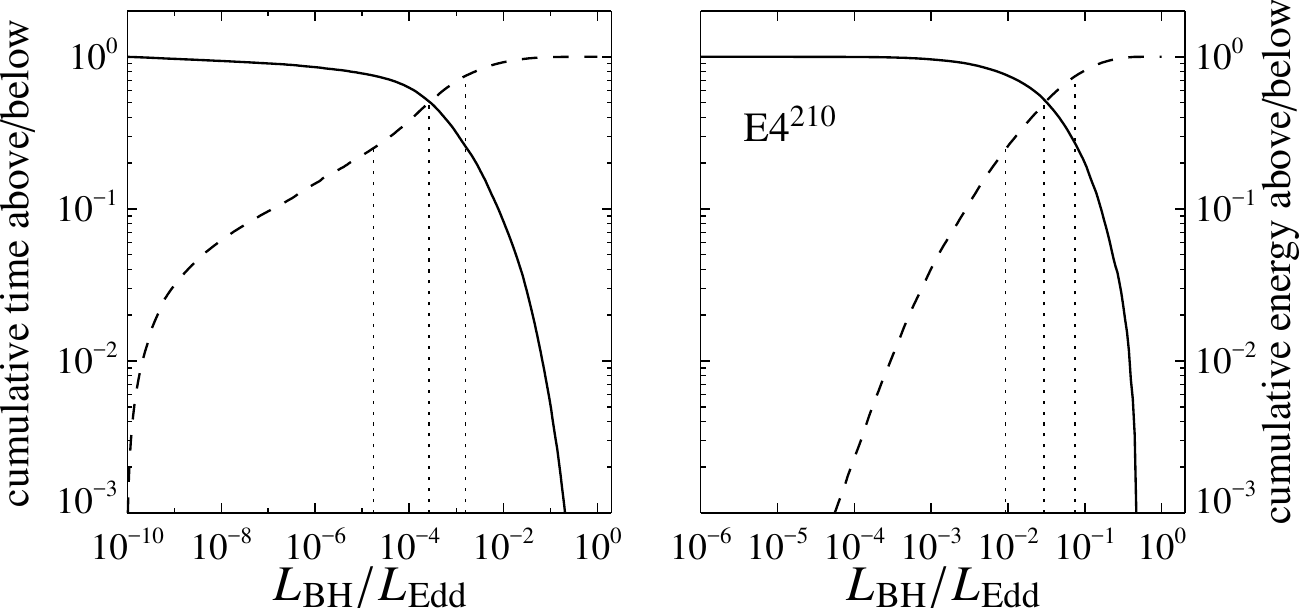}\\

\hskip 2truecm\includegraphics[scale=0.9]{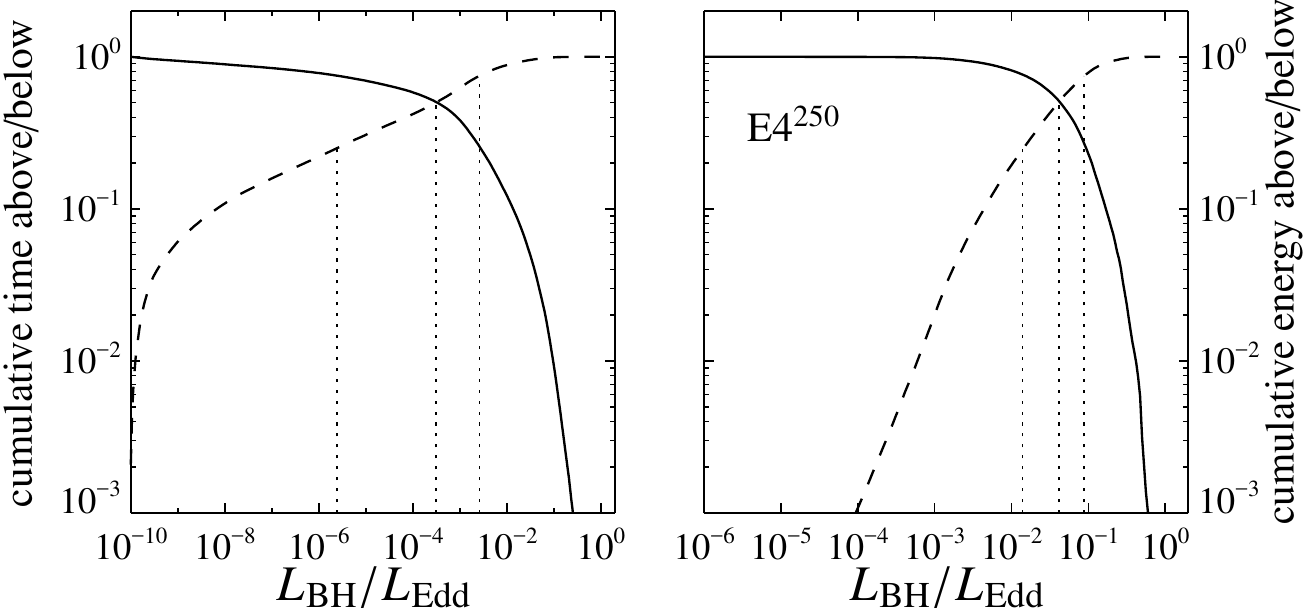}\\

\hskip 2truecm\includegraphics[scale=0.9]{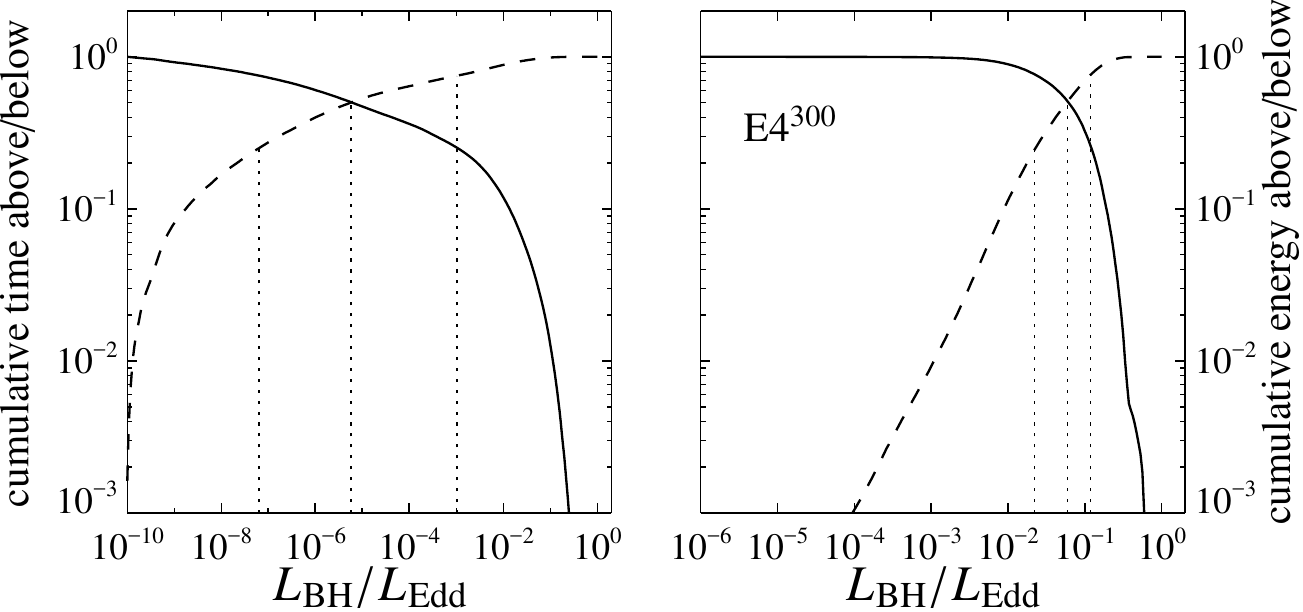}\\
\caption{Left panels: percentage of the total simulation time (11 Gyr)
  spent above (solid line) and below (dashed line) the values of the
  Eddington ratio on the $x$-axis, for the E4 (in black) and E7 (in
  red) FF models.  Right panels: for the same models on the left,
  percentage of the total energy emitted above (solid line) and below
  (dashed line) the Eddington ratio on the x-axis.  In each panel, the
  vertical dashed lines mark the Eddington ratios below which the
  model spends 25\%, 50\% and 75\% of the total time (left panels), or
  below which 25\%, 50\% and 75\% of the total energy is emitted
  (right panels).}
\label{duty4} 
\end{figure*}

\renewcommand\arraystretch{1.4}
\begin{table*}
\centering
\caption{Structural parameters of the galaxy models.}
\vspace{1mm}
\begin{tabular}{ccccccccc}
\toprule
Name         & $\Lb$           & $\re$ & $\Mstar$            & $\Mh$            & $\Mbhz$       &$\se$  & $\fDM $ &  $c$   \\ 
             & $(10^{11}L_{\rm B,\odot})$ & (kpc) & $(10^{11}\Msun)$ & $(10^{11}\Msun)$ & $(10^{8}\Msun)$  &$(\kms)$           &         &        \\
 (1)         &   (2)           &  (3)  &    (4)           &    (5)           &    (6)           &    (7)            &  (8)    &  (9)   \\ 
 \midrule                                                                                                                                
E4$^{180}$   &   0.18          & 3.26  &    0.81          &   16.20          &    0.81          &    160            & 0.62    &   41   \\
E7$^{180}$   &   0.18          & 3.26  &    0.81          &   16.20          &    0.81          &    137            & 0.73    &   41   \\
\midrule                                                                                        
E4$^{210}$   &   0.32          & 4.57  &    1.54          &   30.80          &    1.54          &    187            & 0.62    &   35   \\
E7$^{210}$   &   0.32          & 4.57  &    1.54          &   30.80          &    1.54          &    155            & 0.66    &   35   \\
\midrule                                                                                        
E4$^{250}$   &   0.65          & 7.04  &    3.35          &   67.00          &    3.35          &    223            & 0.63    &   28   \\
E7$^{250}$   &   0.65          & 7.04  &    3.35          &   67.00          &    3.35          &    184            & 0.67    &   28   \\
\midrule                                                                                                                                  
E4$^{300}$   &   1.38          & 11.8  &    7.80          &   160.00         &    7.80          &    267            & 0.66    &   22   \\
E7$^{300}$   &   1.38          & 11.8  &    7.80          &   160.00         &    7.80          &    221            & 0.68    &   22   \\
\bottomrule
\end{tabular}
\flushleft

Notes: $(1)$ Model name: the letters and numbers identify the shape (E4 or E7), the superscript gives the 
$\se$ of the progenitor. $(2)$ Luminosity in the $B$-band. $(3)$ Effective radius (edge-on
view). $(4)$ Total stellar mass. $(5)$ Total DM mass. $(6)$ Black hole initial mass. $(7)$ 
Luminosity-weighted stellar velocity dispersion within a circular aperture of radius $\re/8$, for an edge-on view. $(8)$ Ratio of the
DM mass to the total mass enclosed within a sphere of radius
$\re$. $(9)$ Concentration parameter of the NFW halo. See Sect. 2 for
more details.
\label{tab:params}
\end{table*}
\renewcommand\arraystretch{1.}

\begin{table*}
\caption{Simulations results.}
\vspace{2mm}

\small 
\begin{tabular}{lllllllllllllllllll}
\toprule
name     & $\DMbh$  & ${\DMbh\over\Mbhz}$ & $l_{0.5}$  &$\mathcal{D}$&  $t_L$     & $\DMstar$   &      SFR  &  $t_M$  &  $r_M$         & $\Lx$ & $\Tx$\\
        & ($10^8\Msun$)&  --    & --   &  ($\%$)  &  (Gyr)     & ($10^8\Msun$) & ($\Msun$yr$^{-1}$) & (Gyr) & (kpc)&           ($10^{40}\ergs$)  & (keV)\\
(1)       & (2)        &    (3)     &     (4)    &  (5)       & (6)        &      (7)   &      (8)   &       (9)  & (10)      &     (11)     & (12) \\
\midrule
Rad+mech&         &     &        &   &     &     &      \\  
(FF)&         &     &        &   &     &     &      \\  
\midrule
E4$^{180}$&       0.15 &       0.18 &     0.12 &       5.11 &       4.56 &      14.51 &       0.04 &       5.38 &       9.09 &        0.02     & 0.39  \\
E7$^{180}$&       0.06 &       0.07 &  0.12 &       4.73 &       3.27 &       7.21 &       0.01 &       4.66 &      10.23 &        0.007    & 0.38  \\
E4$^{210}$&       1.54 &       1.01 & 0.03 &       1.60 &       5.04 &      77.60 &       0.28 &       5.25 &       1.68 &        1.27     & 0.47  \\
E7$^{210}$&       1.06 &       0.69 &  0.03 &       1.17 &       4.83 &      64.82 &       0.13 &       5.22 &       1.68 &        0.56     & 0.46  \\
E4$^{250}$&       6.31 &       1.90 &  0.04 &       2.94 &       5.08 &     170.16 &       0.31 &       4.97 &       2.44 &        3.52     & 0.76  \\
E7$^{250}$&       4.91 &       1.48 &       0.04 &       2.18 &       5.24 &     155.71 &       0.44 &       5.24 &       2.29 &        2.76     & 0.63  \\
E4$^{300}$&      23.19 &       3.00 &       0.06 &       3.43 &       5.92 &     342.59 &       1.02 &       5.04 &       4.22 &       21.85   & 1.02  \\
E7$^{300}$&      20.33 &       2.63 &      0.05 &       3.03 &       5.91 &     309.85 &       2.95 &       5.16 &       4.76 &       42.58    & 0.88  \\
\midrule
Mech&      &     &    &        &   &        &       \\  
(MF)&      &     &    &        &   &        &       \\  
\midrule
E4$^{180}$&       0.16 &       0.20 &       --   &      --    &      --    &      14.05 &       0.02 &       5.19 &      11.51 &       0.007     & 0.38        \\
E7$^{180}$&       0.06 &       0.07 &        --   &    --    &      --    &       5.63 &       0.01 &       4.17 &       9.09 &       0.003       & 0.38     \\
E4$^{210}$&       1.80 &       1.18 &       --   &     --    &      --    &      68.44 &       0.33 &       5.83 &       2.93 &       1.47        & 0.48     \\
E7$^{210}$&       1.37 &       0.90 &      --   &      --    &      --    &      61.55 &       0.26 &       5.42 &       2.59 &       1.03        & 0.44  \\
E4$^{250}$&       7.29 &       2.20 &      --   &      --    &      --    &     152.91 &       2.30 &       5.68 &       3.97 &      10.62      & 1.00    \\
E7$^{250}$&       5.74 &       1.73 &      --   &      --    &      --    &     136.96 &       0.88 &       5.91 &       3.74 &       4.39       & 0.64      \\
E4$^{300}$&      22.98 &       2.98 &      --   &      --    &      --    &     278.46 &       0.76 &       5.46 &       6.81 &      18.02      & 1.08     \\
E7$^{300}$&      20.96 &       2.71 &       --   &     --    &      --    &     267.40 &       0.50 &       6.14 &       6.81 &      11.03      & 1.03     \\
\midrule
None       &       &     &    &        &      &        &   \\  
(NOF)       &       &     &    &        &      &        &   \\  
\midrule
E4$^{180}$&       7.82 &       9.76 &       --   &   --    &      --    &      23.59 &       0.25 &       5.35 &       1.47 &       0.30   & 0.31  \\
E7$^{180}$&       4.83 &       6.03 &         --   & --   &      --    &      15.36 &       0.01 &       4.43 &       1.66 &       0.005   & 0.38   \\
E4$^{210}$&      28.17 &      18.47 &         --   & --    &      --    &      53.94 &       0.26 &       5.69 &       0.43 &       1.18   & 0.49  \\
E7$^{210}$&      18.30 &      12.00 &      --   &   --    &      --    &      46.45 &       0.28 &       5.41 &       0.57 &       0.78    & 0.45  \\
E4$^{250}$&      84.20 &      25.38 &        --   & --    &      --    &     115.42 &       0.53 &       5.71 &       0.43 &       4.41   & 0.67  \\
E7$^{250}$&      69.65 &      21.00 &       --   &  --    &      --    &     103.21 &       0.46 &       5.60 &       0.46 &       3.40    & 0.64  \\
E4$^{300}$&     221.23 &      28.66 &       --   &  --    &      --    &     245.76 &       1.24 &       5.88 &       0.50 &      18.75    & 0.94 \\
E7$^{300}$&     195.85 &      25.37 &          --   &  --    &      --    &     224.55 &       1.09 &       5.92 &       0.53 &      15.74   & 0.92 \\
\bottomrule 
\end{tabular} 
\parbox{0.9\linewidth}{\footnotesize %
\vspace{2mm}

\textbf{Notes}. (1) Name of the galaxy model, following the nomenclature of Tab. 1. (2) Total MBH accreted mass. (3)
Percent variation of the MBH mass, with respect to the inital MBH mass
$\Mbhz$. (4) Value of the Eddington ratio $l = \Lbh/\Ledd$ with respect to which the MBH energy is emitted equally above and
below $l$. (5) Duty cycle ($\mathcal{D}$) defined as the ratio between
the time spent by the MBH at $l>0.05$ and the total simulation
time. (6) Time at which half of the total MBH radiation energy, emitted over 2--13 Gyr,
has been emitted, measured since the birth of the galaxy ($t=0$). (7-8) Total mass of stars produced and SFR at 13~Gyr. (9)
Time at which half of the new stellar mass, produced over the simulation time-lapse of 2--13 Gyr, has been created, measured since the birth of the original stellar 
population. (10) Radius containing half of the stars produced by
$t=13$ Gyr. (11) ISM luminosity within 5$\re$ in the 0.3-8 keV band. (12) Emission-weighted temperature calculated as in eq. (43).}

\end{table*}

\end{document}